\documentclass[journal]{IEEEtran}

\usepackage{xcolor}
\usepackage{amsfonts}
\usepackage{algorithmic}
\usepackage{graphicx}
\usepackage{textcomp}
\usepackage{subfigure}
\usepackage{tabularx}
\usepackage{multirow,booktabs}
\usepackage{threeparttable}
\usepackage[ruled,linesnumbered]{algorithm2e}
\usepackage{graphicx}
\usepackage{hyperref}
\hypersetup{
    colorlinks=true,
    linkcolor=black,
    urlcolor=blue     
}
\usepackage[final]{changes}
\usepackage{amsmath, amssymb}

\definecolor{gray}{RGB}{127,127,127}
\definecolor{black}{RGB}{0,0,9}

\newcommand{\graphnumbercolor}[2]{
    \begin{tikzpicture}[baseline=(number.base)]
        \node[fill = #1, circle, inner sep=1.5pt, text=white, outer sep=0pt] (number) at (0, 0, 0) {\footnotesize{#2}};
    \end{tikzpicture}
}

\DeclareMathOperator*{\argmin}{arg\,min}
\begin{document}

\title{Mitigating scalability challenges in LUT-based neural networks via pruning optimisations}

\author{\IEEEauthorblockN{Xuqi Zhu, Huaizhi Zhang, JunKyu Lee, Jiacheng Zhu, Chandrajit Pal, Sangeet Saha, \\ Klaus D. McDonald-Maier and Xiaojun Zhai}
\thanks{All authors are with the School
of Computer Science and Electronic Engineering, University of Essex, Colchester CO4 3SQ, United Kingdom. E-mails: \{xz18173, hz24245, j.lee, jz23222, chandrajit.pal, sangeet.saha, kdm, xzhai\}@essex.ac.uk
}}

\maketitle

\begin{abstract}
Modern deep neural networks heavily rely on a large number of multiply-accumulate operations, which constitute the predominant computational cost. To address this, Look-Up Table (LUT)-based matrix multiplications have emerged as a promising alternative for reducing the computational cost and time of the multiply-accumulate operations in a neural network. However, the LUT-based neural network still faces the scalability challenge due to the inherent limitations of LUT-based matrix multiplication. To mitigate these scalability limitations, this paper proposes a scalable and energy-efficient LUT-based approximate matrix multiplication unit (LUT-MU) constituting the basic component of the neural networks by integrating a pruning strategy on the MADDNESS algorithm, a LUT-based matrix multiplication methodology. With increasing problem size and precision demands in matrix multiplication, our proposed LUT-MU architecture effectively constrains resource expansion. The case study shows that deploying our LUT-MU in neural network architectures, including fully connected layers (MNIST) and ResNets (CIFAR-10, ImageNet)—on XCZU7EV and XCZU19EG FPGAs, 
\deleted{produces up to $2 \times$  throughput improvement and $10 \times$  energy efficiency gains over leading quantised neural network implementations, with moderate impact on accuracy.}
\added{produces up to $1.6 \times$ throughput improvement and $4.2 \times$ energy efficiency gains over mainstream CUDA-based network implementations, and $1.8\times$ energy efficiency compared to leading quantised neural network implementations, with moderate impact on accuracy.}
Compared to original MADDNESS-based neural networks, our LUT-MU shows $1.3$ to $2.6\times$ resource savings based on various resolution configuration settings of MADDNESS. 
\end{abstract}


\IEEEpeerreviewmaketitle

\section{Introduction}

Recent years have seen notable progress in the field of hardware-based inference accelerators that implement Neural Networks (NNs) for AI \cite{hu2022survey, samanta2024survey}. As such, AI application scenarios continue to expand, and it becomes a new challenge to satisfy various performance-oriented requirements beyond accuracy in various tasks \cite{gao2023application, zhu2023bayesian, gao2023modelling}. \added{For example, consider applications such as person or vehicle tracking with fixed camera angles, lane detection in good visibility environments deployed on energy-strained autonomous vehicles, or a simple object recognition task of a robot working on well-illuminated indoor environments. In these scenarios, where resources are constrained, and energy efficiency is critical but the system is insensitive to accuracy degradation, trading an moderate reduction in accuracy for significant improvements in throughput and energy efficiency is feasible and desirable \cite{boudjadar2025dynamic, mounesan2025infer, fan2023sparse}. }
To handle the diversity of AI applications and the sophistication of task scenarios, researchers are exploring the design simplification of resource-efficient high-throughput multipliers in NN accelerators that execute AI applications at different problem scales and performance requirements \cite{lee2023resource, hamanaka2023exploration, guo2024accelerating}. Since matrix multiplication constitutes a major computational overhead in NN inference tasks, an optimised matrix multiplication unit can improve the overall accelerator performance. 


\deleted{To address computational resource constraints, researchers have proposed product quantisation algorithms \cite{PQtech, PQtech2}, which enable approximate matrix multiplication by leveraging Look-Up Tables (LUTs), known as LUT-based approximate matrix multiplication. LUT-based approximate matrix multiplication algorithms are well-suited for tasks such as accelerating dense linear layers, kernel-based classification, and image filtering if one of the two vector elements is known. Recent studies have demonstrated its effectiveness on the CIFAR-10 ImageNet dataset, showing that LUT-based approximate matrix multiplication can replace entire middle layers of neural networks, not just the final dense layer, while still maintaining competitive accuracy~\cite{halutmatmul}. Various LUT-based approximate matrix multiplication algorithms were studied to accelerate neural networks on platforms such as CPUs, GPUs, ASICs, FPGAs, and Computing in Memory (CIM) \cite{LUTIN, wang2025evasion, halutmatmul, PQA, peku}, showing significant resource savings, compared to existing matrix multiplication methodologies. }


\deleted{The deployment of LUT-based neural networks remains challenging due to the scalability issues inherent in its LUT-based matrix multiplication approaches \cite{LUTMAC, CompressedLUT}. These scalability issues are further worsened by redundant computations across sequential layers, including both intra-layer and inter-layer redundancy (i.e., parameters in each layer and data movement between layers). However, we note that the index-based clustering mechanism of product quantisation \cite{Blalock2021} can restrict dot-product computations in successive layers to selected indices, offering a natural opportunity for computational pruning in neural networks. The LUT-based approximate matrix multiplication also exhibits computation patterns different to traditional matrix multiplication operations, resulting in data dependency and random memory access patterns in computing neural network inference. Naively applying pruning without considering these atypical computation patterns poses challenges for executing LUT-based approximate matrix multiplication efficiently on hardware.}

\added{The quantisation is one of the most widely adopted techniques for reducing multiplication complexity and memory requirements \cite{Liang2018, Blott2018, Umuroglu2017}. In parallel, researchers have explored matmul-free techniques, such as product quantisation \cite{PQtech, PQtech2} and polynomial expression \cite{polylut}, as alternatives to computation-intensive matrix multiplication. However, the deployment of LUT-based neural networks remains challenging due to the scalability issues inherent in their LUT-based matrix multiplication approaches \cite{LUTMAC, CompressedLUT}. These scalability issues are further worsened by redundant computations across sequential layers, including both intra-layer and inter-layer redundancy (i.e., parameters in each layer and data movement between layers). However, we note that the index-based clustering mechanism of product quantisation \cite{Blalock2021} can restrict dot-product computations in successive layers to selected indices, offering a natural opportunity for computational pruning in neural networks. Additionally, the LUT-based approximate matrix multiplication also exhibits computation patterns different to traditional matrix multiplication operations, resulting in data dependency and random memory access patterns in computing neural network inference. Naively applying pruning without considering these atypical computation patterns poses challenges for executing LUT-based approximate matrix multiplication efficiently on hardware.}


 To address this, we propose an LUT-based Approximate Matrix Multiplication Unit (LUT-MU) that \deleted{overcomes} \added{mitigates} the inherent scalability limitations of existing LUT-based matrix multiplication methodologies. The LUT-MU built on MADDNESS with a tailored pruning strategy allows LUT-based NNs to achieve a significant reduction in memory footprint and resource consumption growth with problem size and precision demands. The proposed LUT-MU with tailored pruning strategy through hardware-software co-design also allows NNs to achieve significant gains in throughput and energy efficiency, compared to typical element-wise quantised matrix multiplications.

The contributions of this paper are summarised as follows:

\begin{itemize}
\item We first propose pruning optimisation to eliminate inter-layer (data movement) and intra-layer (parameter) redundancies of MADDNESS matrix multiplications used in NN inferences to \deleted{address} \added{migrate} scalability issues from LUT-based neural networks, \added{thereby expanding design space within the trade-off between performance and resource constraint}.
\item By eliminating redundancies, we reduce bandwidth consumption and mitigate resource overhead growth as the matrix multiplication problem size increases, thereby significantly improving the energy efficiency and throughput of LUT-based approximate matrix multiplication.
\item We explore the hardware design space and propose customised memory allocation and access design to optimise the data dependency and incoherent memory access pattern in LUT-based approximate matrix multiplication.
\item \deleted{Our case study shows that LUT-MU neural networks such as fully connect layers (SFC) and ResNets achieve up to $2 \times$ throughput and $10 \times$ energy efficiency (GOPS/W) on Xilinx XCZU7EV and XCZU19EG FPGAs over state-of-the-art QNNs with minor accuracy loss.}
\added{Our case study shows that LUT-MU neural networks such as ResNets achieve up to $1.6\times$ higher throughput and $4.2 \times$ energy efficiency (GOPS/W) on Xilinx XCZU7EV and XCZU19EG FPGAs over leading QNN implementation with moderate accuracy loss.}
\end{itemize}


\deleted{In the following sections, we highlight the existing studies in Section~\ref{section:related_work}, followed by a discussion of MADDNESS 
in Section~\ref{section:background}. Section~\ref{sec:challenges} analyses the challenges of applying LUT-based approximate matrix multiplication on hardware. The general architecture and design of the proposed LUT-MU are presented in Section~\ref{section:AMU}. Section~\ref{section:result} compares the performance of our proposed LUT-MU compared to the previous works. Finally, Section~\ref{section:conclusion} concludes along with the future challenges.}

\section{Related works}\label{section:related_work}


Several studies used quantised matrices to simplify multiplication computation and memory requirements. For example, the researchers in \cite{Liang2018, Blott2018, Umuroglu2017} utilised Binarised Neural Networks (BNN) to save computational resources and improve throughput by substituting matrix multiplications with XNOR and pop-count operations. Subsequently, Wang et al. \cite{Wang2020} and Zhang et al. \cite{Zhang2021} further optimised the performance of BNN from the aspect of enhancing the logic density of XNOR gate-based matrix multiplication and mitigating accuracy degradation caused by low-precision quantisation.

Moreover, some researchers proposed LUT-based approximate matrix multiplication by using product quantisation \cite{PQtech, PQtech2}. \added{Various studies across platforms such as CPUs, GPUs, ASICs, FPGAs, and Computing in Memory (CIM) \cite{LUTIN, wang2025evasion, halutmatmul, PQA, peku}, showing that product quantisation delivers significant performance gains over conventional matrix-multiplication techniques.} Among these, we note that the MADDNESS algorithm is one of the most aggressive approaches. The MADDNESS~\cite{Blalock2021} streamlines the processing by eliminating the need for online multiplication, which is attractive in a scenario that requires a large number of matrix multiplications with known weight matrices, especially in deep learning models. However, Tang et al. \cite{Tang2023} noticed that LUT-based approximate matrix multiplication, such as MADDNESS, is nonderivable, hence it ignores the loss from backward, resulting in a significant accuracy drop when multilayers are replaced by LUT-based approximate matrix multiplication. To restore the inference accuracy, Jannis et al. \cite{halutmatmul} presented a linear algebra arithmetic called `Halutmatmul' to substitute the non-derivable hash functions involved in `Encode', which allows NNs using MADDNESS can be retrained by backpropagation. In contrast, Fuketa et al. \cite{fuketa} employed additional LUTs to reduce the loss of accuracy by introducing residual product quantisation. Although researchers \cite{LUTIN, tabconv} have reported the attractive capability of the LUT-based approximate matrix multiplication algorithm, 
research in \cite{zhu2025late} indicates that the general processor architectures enhanced by well-designed FPU are not well geared to the multiply-accumulate (MAC) free computation. This is due to the distinct computational patterns and random memory access behaviours inherent to LUT-based approximate matrix multiplication, which differ significantly from traditional matrix multiplication operations. These limitations motivate the exploration of efficient hardware architectures tailored to such computations. Although some studies \cite{halutmatmul, fuketa, zhu2025late} have proposed dedicated architectural designs to address the throughput bottlenecks caused by the distinct computational patterns and random memory access behaviours of LUT-based matrix multiplication, these approaches still face scalability challenges inherent to LUT-based matrix multiplication. Consequently, we devised a specialised architectural design incorporating dedicated pruning optimisation on MADDNESS, thereby mitigating scalability challenges in LUT-based approximate matrix multiplication.

\section{Scalability Challenge}

\begin{figure}[htb]
    \centering
    \includegraphics[width=\linewidth]{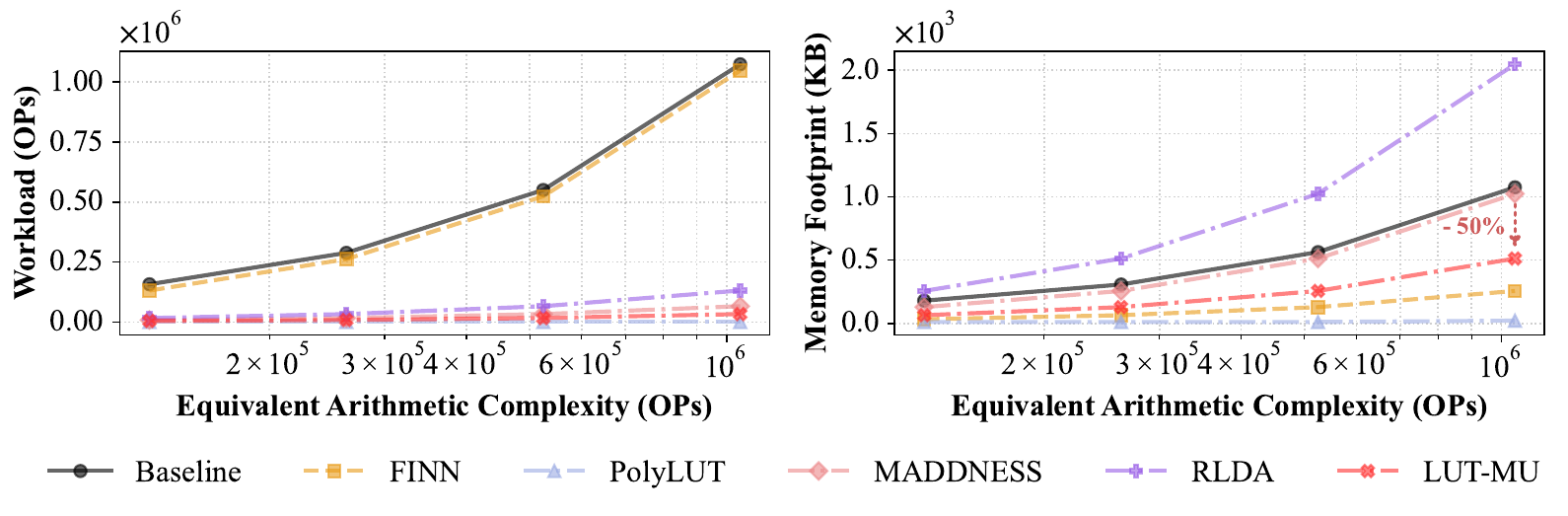}
    \caption{\added{Actual arithmetic workload (left) and memory footprint (right) as the equivalent arithmetic complexity of matrix multiplication increases, across the exact (baseline), quantised (FINN \cite{Blott2018}), LUT-based approximated (MADDNESS \cite{Blalock2021}, RLDA \cite{fuketa}, LUT-MU) matrix multiplication and polynomial operators (PolyLUT \cite{polylut}). }}
    \label{fig:motivation}
\end{figure}

\added{The quantisation is one of the most widely adopted techniques for reducing multiplication complexity and memory requirements \cite{Liang2018, Blott2018, Umuroglu2017}. In parallel, researchers have explored matmul-free techniques, such as product quantisation \cite{PQtech, PQtech2} and polynomial expression \cite{polylut}, as alternatives to computation-intensive matrix multiplication. As illustrated in Fig. \ref{fig:motivation}, instead of using linear operators (matrix multiplication), PolyLUT \cite{polylut} significantly reduces both time and space complexity at an equivalent complexity by storing pre-computed polynomial operator results in LUTs rather than using linear operators. However, the polynomial operators impede the training of more complex neural networks using backpropagation algorithms, limiting their applicability to deeper NN models and sophisticated tasks. Meanwhile, LUT-based approximated matrix multiplications using product quantisation (e.g., MADDNESS \cite{Blalock2021}, RLDA \cite{fuketa}, LUT-MU) eliminate element-wise computation in exact (baseline) and quantised (FINN \cite{Blott2018}) matrix multiplication, thereby achieving equivalent arithmetic complexity while requiring substantially less computation workload. }

\added{However, the deployment of LUT-based neural networks remains challenging due to the scalability issues inherent in its LUT-based matrix multiplication approaches \cite{LUTMAC, CompressedLUT}. As shown in Fig. \ref{fig:motivation}, the pre-computed LUT (MADDNESS, RLDA) introduces additional memory footprint compared to quantised matrix multiplication (FINN), creating a dilemma for efficient hardware design as model size grows: designers either allocate more resources to sustain performance or accept performance degradation to conserve resources. These scalability issues are further worsened by redundant computations across sequential layers, including both intra-layer and inter-layer redundancy (i.e., parameters in each layer and data movement between layers).}

\added{To address this, we propose LUT-MU that mitigates the inherent scalability limitations of existing LUT-based matrix multiplication methodologies. By leveraging the natural property of the index-based clustering mechanism of product quantisation, the LUT-MU with a tailored pruning strategy allows LUT-based NNs to achieve a 50\% reduction in memory footprint and resource consumption growth with arithmetic complexity expansion compared to previous works (MADDNESS) in the same configuration setting ($d_{sub} = 8$, $I = 4$). Together with hardware-software co-design, the proposed LUT-MU with tailored pruning strategy also allows NNs to achieve significant gains in throughput and energy efficiency, compared to typical element-wise quantised matrix multiplications. }

\section{Background: product quantisation}\label{section:background}

To help understand the proposed pruning operation and corresponding hardware design, we first discuss the workflow of product quantisation, taking MADDNESS as an example. The MADDNESS is an LUT-based approximate multiplication algorithm, derived from the Product Quantisation (PQ) technique \cite{PQtech, PQtech2} which organises datasets into a fixed number of clusters and learn to substitute for sample vectors within each cluster. The MADDNESS requires offline training to generate an LUT for the online multiplication phase.


\subsection{Offline training}\label{sec:offline}

We refer to ``offline training" in this paper as the process used to compute partial dot product values in LUTs, distinct from the typical training method involving backpropagation used for training neural networks. In the first stage of the offline training, MADDNESS generates codebooks. Each codebook consists of a set of prototypes learned from the training data $\tilde{A} \in \mathbb{R}^{N\times D}$.  For example, using the $i$-th training sample, $\tilde{A}_i \in \mathbb{R}^{D}$ can be divided into $C$ $\mathbb{R}^{D/C}$ dimensional data, where $C$ is an integer that allows ${D/C} = d_{sub}$ to be an integer. In this case, there exist $C$ codebooks according to $C$ subdimensions of training data $\tilde{A}$. We use the notation $\tilde{A}^{(c)}_i$ for the $c^{th}$ sub-dimensional vector in the $i$-th training sample. \deleted{ As shown in Fig.~\ref{fig:AMM}. (a), the 1st subdimensional vectors formed the training set $\tilde{A}^{(1)}$, which is used to generate the $1^{st}$ codebook. To simplify the formulation, we use $codebook_{i}$ to denote the $i^{th}$ subdimensional vector of the given input vector, since each subdimensional vector can only be represented by a specific prototype from a certain codebook.} The offline training can be summarised as the following equations \cite{Blalock2021}:
\begin{align}
    & \sum_{i=1}^{N}\sum_{c=1}^{C}\sum_{j\in{\mathcal{J}^{(c)}}} (\mathbf{\tilde{A}}^{(c)}_{i,j} - \mathbf{P}^{(c)}_{\mathbf{z}_i^{(c)},j})^2\label{math:prototypes}\\
    &  lut^c_g(\mathbf{b}) \triangleq \sum_{j\in{\mathcal{J}^{(c)}}}\mathbf{b}_j^{(c)}  \mathbf{P}^{(c)}_{g,j},~ 0 < g\leq G \label{math:lut}
\end{align}
To simplify without loss of generality, let us denote $\mathbf{P}^{(c)} \in \mathbb{R}^{G \times|\mathcal{J}^{(c)}|}$ is the codebook prototypes $c$, where $G$ is the number of prototypes for each codebook, and $|\mathcal{J}^{(c)}|$ is the length of the index set associated with codebook $c$. In the offline training stage, MADDNESS uses a heuristic clustering strategy based on a $I$-level decision tree (i.e., $I$-rounds bisection) to find $G = 2^I$ prototypes $\mathbf{P}^{(c)}$ for each codebook $c$, so that Equation~(\ref{math:prototypes}) is minimized \cite{Blalock2021}. Here, $\mathbf{z}^{(c)}\in \mathbb{R}^N$ has $N$ indexes ranged from $(0, G]$, indicated that the $\mathbf{P}^{c}_{\mathbf{z}^{c}}$ are the best approximated to the $c$-th sub-dimensional vectors of all $N$ samples in the dataset $\tilde{A}$ (i.e. $\tilde{A}^{(c)}$).

\subsection{Online multiplications (Inference)}\label{sec:maddness_online}

Online multiplications assume that the patterns in the test dataset are similar to those in the training dataset. Under this assumption, we can utilise the following equations to compute the approximate multiplications \cite{Blalock2021}: 
\begin{align}
    & encode^c(\mathbf{a}) \triangleq \argmin_{\{g | 0 < g \leq G\}}\sum_{j\in \mathcal{J}^{(c)}} (\mathbf{a}_j^{(c)} - \mathbf{P}^{(c)}_{g,j})^2 \label{math:encode}\\
    & \mathbf{ab}^T \approx \sum_{c = 1}^{C} lut^c_g(\mathbf{b}),~ g = encode^c(\mathbf{a}) \label{math:aggregate}
\end{align}

\begin{figure}[htb]
    \centering
    \includegraphics[scale = 0.475]{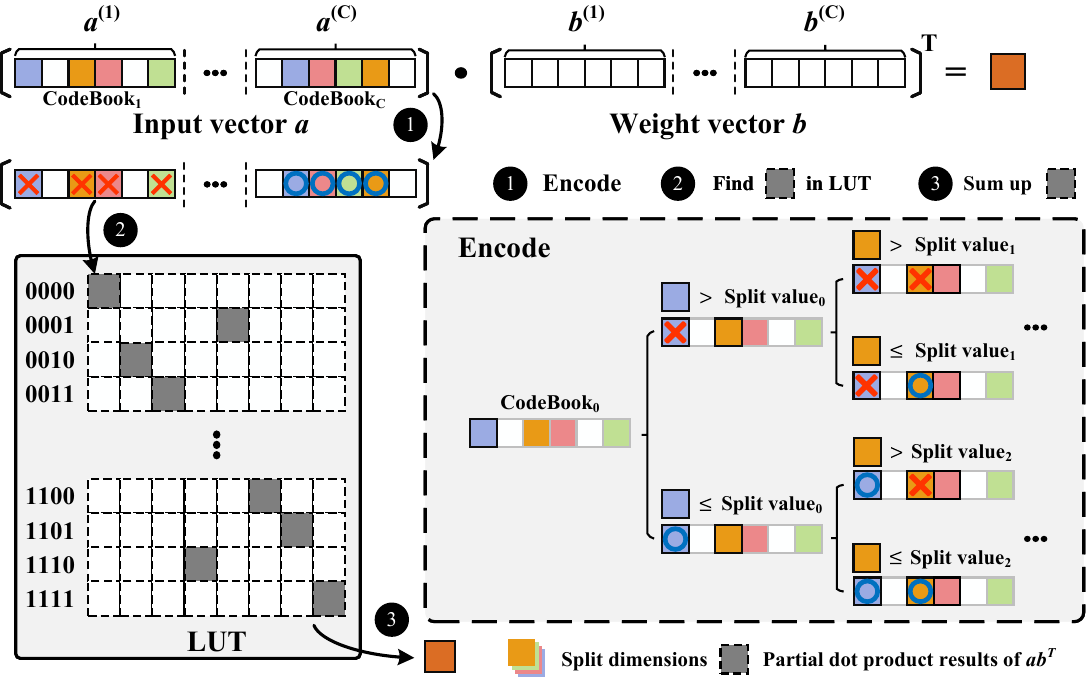}
    \caption{MADDNESS online multiplications.}
    \label{fig:AMM}
\end{figure}

In the online multiplication stage, a test sample (i.e., input vector $\mathbf{a}$) is first divided into $C$ subdimensional vectors. PQ techniques then employ the general `Encode' Equation~(\ref{math:encode}) to identify the prototype from $G$ prototypes that are most similar to the given sub-dimensional vector $\mathbf{a}^{(c)}$. Specifically, the Encode operation in MADDNESS is a binary decision tree search process with a depth of $I$, which sequentially compares $I$ split values with corresponding $I$ values in the split dimensions (i.e. colored cells in codebooks) of $\mathbf{a}^{(c)}$ to find the best matching prototype, as illustrated in Fig.~\ref{fig:AMM}\deleted{. (b)} \graphnumbercolor{black}{1}. Here, the $C\times (2^{I}-1)$ split values in the decision tree and the $C\times I$ split dimensions in the vector $\mathbf{a}$ are obtained during the training stage.

Once the prototypes for each codebook are determined, the approximate partial dot products (grey cells) are retrieved from the LUT as shown in Fig.~\ref{fig:AMM}\deleted{. (b)} \graphnumbercolor{black}{2}. For example, binary decisions derived from the split dimensions of the first subdimensional vector $\mathbf{a}^{(1)}$ are ``XXXX" (equivalent to ``0000"), while those from the last subdimensional vector $\mathbf{a}^{(C)}$ are ``OOOO" (equivalent to ``1111"), serving as the LUT address for a partial dot product value. Finally, MADDNESS applies Equation (\ref{math:aggregate}) to compute approximately $\mathbf{ab}^T$ by aggregating all grey cells in LUT, as depicted in Fig.~\ref{fig:AMM}\deleted{. (b)} \graphnumbercolor{black}{3}.

\subsection{Challenges posed by MADDNESS}\label{sec:challenges}

Previous research has progressively improved MADDNESS to enable its use as a replacement for exact matrix multiplication in deep learning models. 
However, MADDNESS and its derivative LUT-based matrix multiplication methods require storing a large number of precomputed results offline to approximate exact matrix multiplication, which leads to scalability challenges as the matrix multiplication problem size and accuracy requirements increase. Moreover, modern deep learning models frequently involve cascades of matrix multiplications. In such scenarios, simply substituting matrix multiplications with MADDNESS causes unnecessary data movements and parameter storage. In addition, the atypical computation pattern of MADDNESS introduces data dependency and incoherent memory access behaviours. These redundancies and atypical computation patterns exacerbate the consumption of effective bandwidth and memory resources, which ultimately restricts computational throughput and energy efficiency as computational complexity and accuracy requirements increase.

\begin{figure}[htb]
    \centering
    \includegraphics[scale = 0.52]{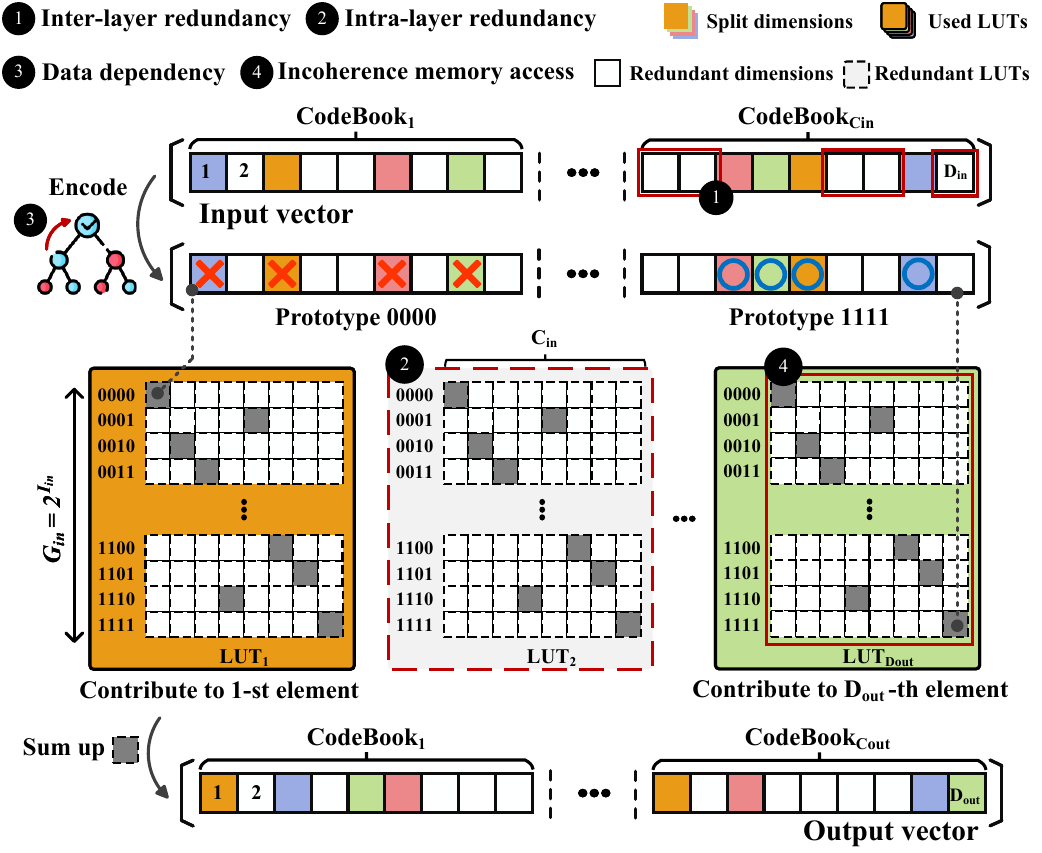}
    \caption{Example of computing continuous LUT-based matrix multiplication in a NN. This example illustrates a scalability challenge in MADDNESS is raised from two key factors: 1. inter-layer redundancies, 2. intra-layer redundancies. In addition, atypical computation pattern introduces 3. incoherent memory access behaviour and 4. data dependency in decision-tree based encoding further restricts computation pipeline throughput.}
    \label{fig:challenge}
\end{figure}

Fig.~\ref{fig:challenge} describes this redundancy issue when substituting matrix multiplications with MADDNESS in a fully connected layer.
For example, the precomputed partial inner products between the input vector prototypes and $D_{out}$ partial weight vectors are stored as $D_{out}$ LUTs. By clustering training data $\tilde{A}$ as described in Section \ref{sec:offline}, these input vectors can share the same split dimensions in the online inference phase. However, according to the `Encode' processing described in Section \ref{sec:maddness_online}, only the values in split dimensions contribute to computing the multiplication, while the values in other dimensions (i.e., the white dimension in codebooks in Fig.\ref{fig:challenge}) delivered to the next computation unit are considered as inter-layer redundancies (Fig.~\ref{fig:challenge}. \graphnumbercolor{black}{1}). Similarly, only colored LUTs contribute to computing values in split dimensions of the next layer, while other LUTs in the current computation unit are considered as intra-layer redundancies (Fig.~\ref{fig:challenge}. \graphnumbercolor{black}{2}). As a result, MADDNESS generates numerous superfluous parameters and transmits massive amounts of redundant data, as shown in Fig. \ref{fig:challenge}. Moreover, these redundancies consume effective bandwidth while occupying substantial storage. The negative impact of these redundancies is exacerbated when they encounter larger matrix multiplication, exacerbating the time and space complexities. From the hardware design perspective, the decision-tree based encoding recursively processes values along split dimensions, introducing data dependency (Fig.~\ref{fig:challenge}. \graphnumbercolor{black}{3}) that hinders parallelism and thereby limits computation throughput. Additionally, for a given input vector, only $1/2^I$ of the partial dot products (i.e., grey cells) in the LUTs contribute to the inference calculations. Upon retrieval, these partial results were scattered randomly across the LUT to obtain the approximated results. This results in incoherent memory access patterns (Fig.~\ref{fig:challenge}. \graphnumbercolor{black}{4}), which further reduce the effective bandwidth and impede the hardware design from achieving its theoretical maximum throughput.

\section{Proposed Methodology: LUT-MU}\label{section:AMU}

\deleted{To enhance effective bandwidth while minimising the impact of increasing the size of the matrix multiplication problem on computational resource and throughput (II) of the LUT-MU, we design LUT-MU which optimise MADDNESS \cite{Blalock2021} by eliminating inter-layer and intra-layer redundancies. Additionally, we introduce a trade-off between memory allocation and access design to overcome the throughput bottleneck caused by the incoherent memory access pattern in MADDNESS. }

\added{To migrate the performance degradation of LUT-based approximated matrix multiplication arising from its inherent scalability limitations, we design LUT-MU, which optimises MADDNESS \cite{Blalock2021} by eliminating inter-layer and intra-layer redundancies. Additionally, we introduce a trade-off between memory allocation and access design to overcome the throughput bottleneck caused by the incoherent memory access pattern in LUT-based approximated matrix multiplication.}


\subsection{Algorithm Optimisation}

\deleted{We apply three design optimisation strategies to eliminate inter-layer and intra-layer redundancies in MADDNESS: 1) data pruning, 2) data reshape, and 3) parameter pruning, which make LUT-MU more efficient and compact. Furthermore, eliminating redundancies can improve the effective bandwidth of LUT-MU and enable it to decouple the time and space complexities from the problem size. This helps minimise the impact of the increasing problem size on resource utilisation and the throughput (II) of the LUT-MU.} 

\added{We apply three design optimisation strategies to eliminate inter-layer and intra-layer redundancies in MADDNESS: 1) data pruning, 2) data reshape, and 3) parameter pruning, which expands the optimisation room of LUT-MU within the dilemma of performance degradation and resource constraint.}

\begin{figure}[htbp]
    \centering
    \includegraphics[scale = 0.6]{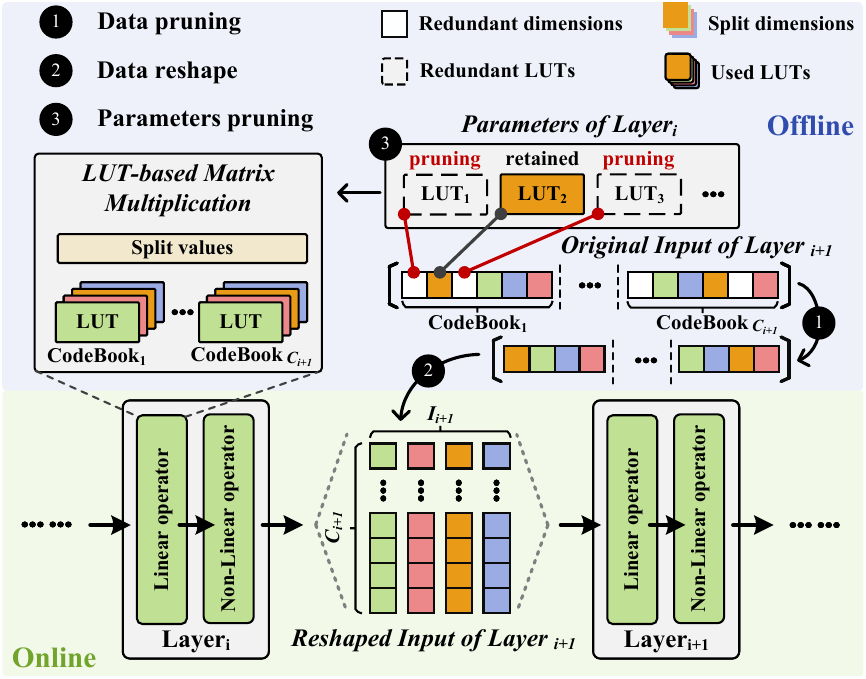}
    \caption{Data pruning and parameter pruning: inter-layer and intra-layer redundancies can be streamlined when multiple LUT-based approximate matrix multiplication units are used sequentially.}
    \label{fig:optimisation}\vspace{-0.5cm}
\end{figure}

\subsubsection{Data pruning}\label{sec:opt_1}



As shown in Fig.~\ref{fig:optimisation}, most NN models are constructed by cascading repeated computation blocks, and the operations in each block consist of linear operators (e.g., matrix multiplication, reshape) and nonlinear operators (e.g. batch normalisation, activation) \cite{Blott2018, peku}. In such scenarios, substituting matrix multiplications with MADDNESS in a layer introduces inter-layer redundancies as discussed in Section~\ref{sec:challenges}. However, all split dimensions in input vectors are predetermined during the offline training stage described in Section \ref{sec:offline}, and \deleted{the most of the nonlinear operation} \added{all operations} in computation blocks cannot hide or change the split dimensions. These properties expose us to an opportunity to eliminate inter-layer redundancies. As illustrated in Fig. \ref{fig:optimisation}.\graphnumbercolor{black}{1}, the data pruning technique retains solely the vital information including the split dimensions and the values at the split dimensions (e.g. coloured blocks in the codebooks as depicted in Fig.~\ref{fig:optimisation}) prior to computing approximate matrix multiplication\deleted{, thus decoupling the time complexities from the problem size and minimising the impact of increasing problem size on throughput (II).} \added{. By pruning redundant LUTs and dimensions, we can alleviate the growth of the memory footprint with increasing problem size, thereby minimising the impact of increasing problem size on resource utilisation and throughput (II).} However, certain non-linear operators (e.g., softmax) require obtaining complete information from all dimensions. \deleted{As a result, our pruning strategy is less effective for the preceding matrix multiplication operator, but subsequent matrix multiplication operations remain unaffected. } \added{As a result, our pruning strategy is less effective for the matrix multiplication operator directly following these operators, while earlier matrix multiplication operations remain unaffected.}

\subsubsection{Data reshape}\label{sec:opt_2}

Data pruning fragments the entire input vector, dispersing valid information (i.e., the values of the split dimensions) randomly across each codebook. This fragmented input vector poses challenges in the design of pipelines in LUT-MU. To overcome this issue, we restructure the pruned input vector. As shown in Fig.~\ref{fig:optimisation}. {\centering \graphnumbercolor{black}{2}}, the values on the split dimensions used in $i$-th round encoding are assembled into the $i$-th cluster, where $i \in [1, I]$ and $I$ is the number of split dimensions. Each cluster comprises $C$ blocks from the $C$ corresponding codebooks, where $C$ is the number of the codebook in a given input. Additionally, the output feature map needs to be restructured to match the input structure of the next LUT-MU to facilitate the cascading of the LUT-MUs.

\subsubsection{Parameters pruning}\label{sec:opt_3}

The original MADDNESS algorithm requires loading all LUTs to compute approximate matrix multiplication between the input and weight matrix to obtain the entire output. These processes introduce intra-layer redundancies when multiple approximate matrix multiplication units are employed consecutively. However, data pruning eliminates the need to load all LUTs. As shown in Fig.~\ref{fig:optimisation}. \graphnumbercolor{black}{3}, for the current computation block (i.e., $Layer_{i}$), only the colored LUTs contribute to computing values in split dimensions of the next computation block (i.e., $Layer_{i+1}$), while other LUTs merely produce redundant IO data. Therefore, the parameter pruning strategy only conserves the LUTs that are responsible for computing vital information (i.e., the value in split dimensions) of the next computation block. By streamlining the parameter footprint, \deleted{LUT-MU can decouple the space complexities from the problem size and minimise the impact of increasing problem size on memory resource utilisation.}
\added{LUT-MU expands design space within the trade-off between performance and resource constraint.}

\subsubsection{Pruning on Convolution}

\begin{figure}[h]
    \centering
    \includegraphics[width = \linewidth]{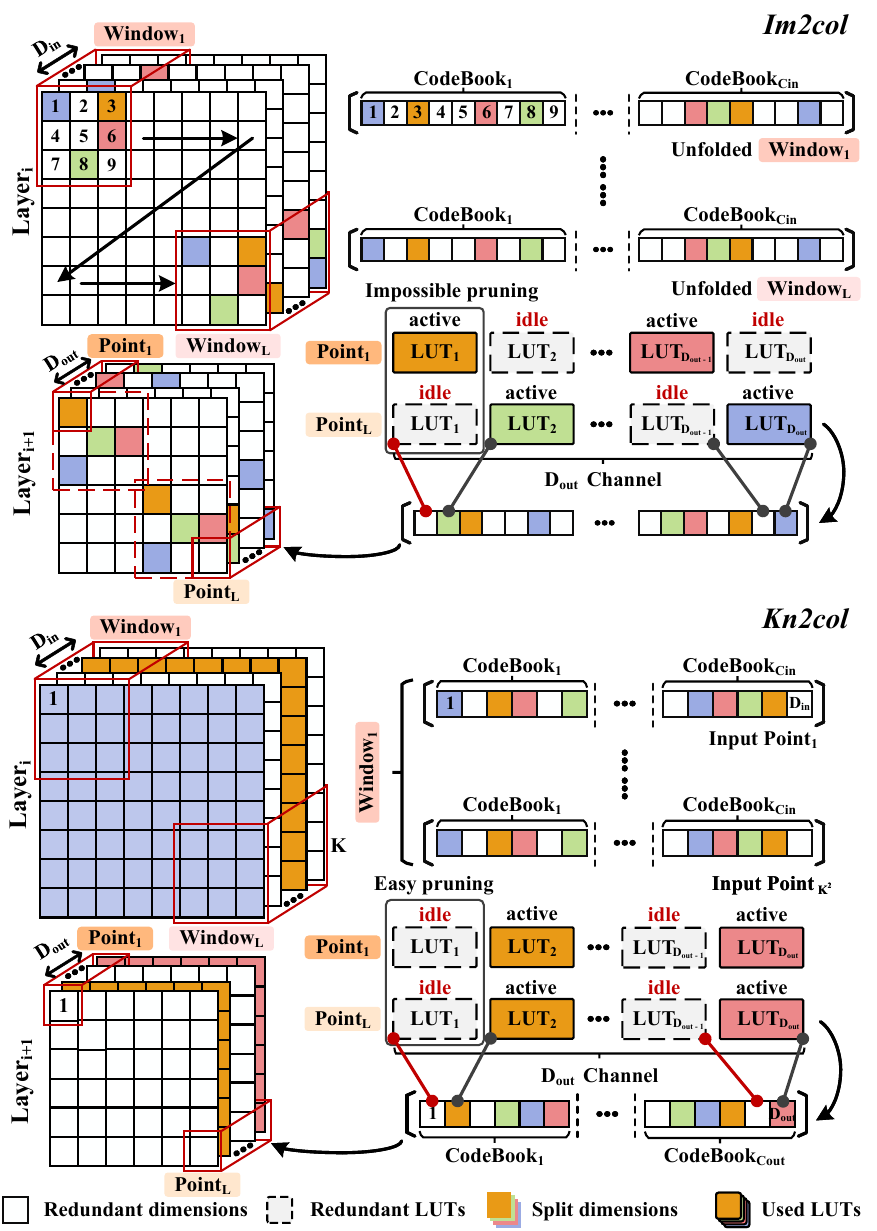}
    \caption{Pruning example in the convolution: Im2col-based convolution impedes effective pruning, while Kn2col-based convolution helps pruning optimisation. }
    \label{fig:kn2col}
\end{figure}


To better understand the pruning optimisation for LUT-based matrix multiplication, let us take the example of convolution computation in CNNs. Previous works \cite{halutmatmul, tabconv} leverage Im2col to convert the convolution operation into multiple matrix multiplications by flattening convolution windows to a series of vectors (i.e., unfolded windows). As illustrated in Fig.~\ref{fig:kn2col}, $(K\times K, D_{in})$ convolution window (e.g., the region of the input feature map enclosed by the red box) is reshaped into an unfolded window comprising $C_{in}$ codebooks, each containing $K \times K$ values. Here, $K$ denotes the convolution kernel width. The convolution operation is finally replaced by $L$ rounds of LUT-based matrix multiplications between unfolded window and $D_{out}$ weight kernels, each computing a $(1 \times 1, D_{out})$ output point. In each round of multiplication, only the active LUTs (i.e., the colored LUTs in Fig.~\ref{fig:kn2col}) contribute to the computation of specific values in the split dimensions of the next layer, while other idle LUTs are treated as redundancies.

\begin{figure*}[h]
    \centering
    \includegraphics[width = \linewidth]{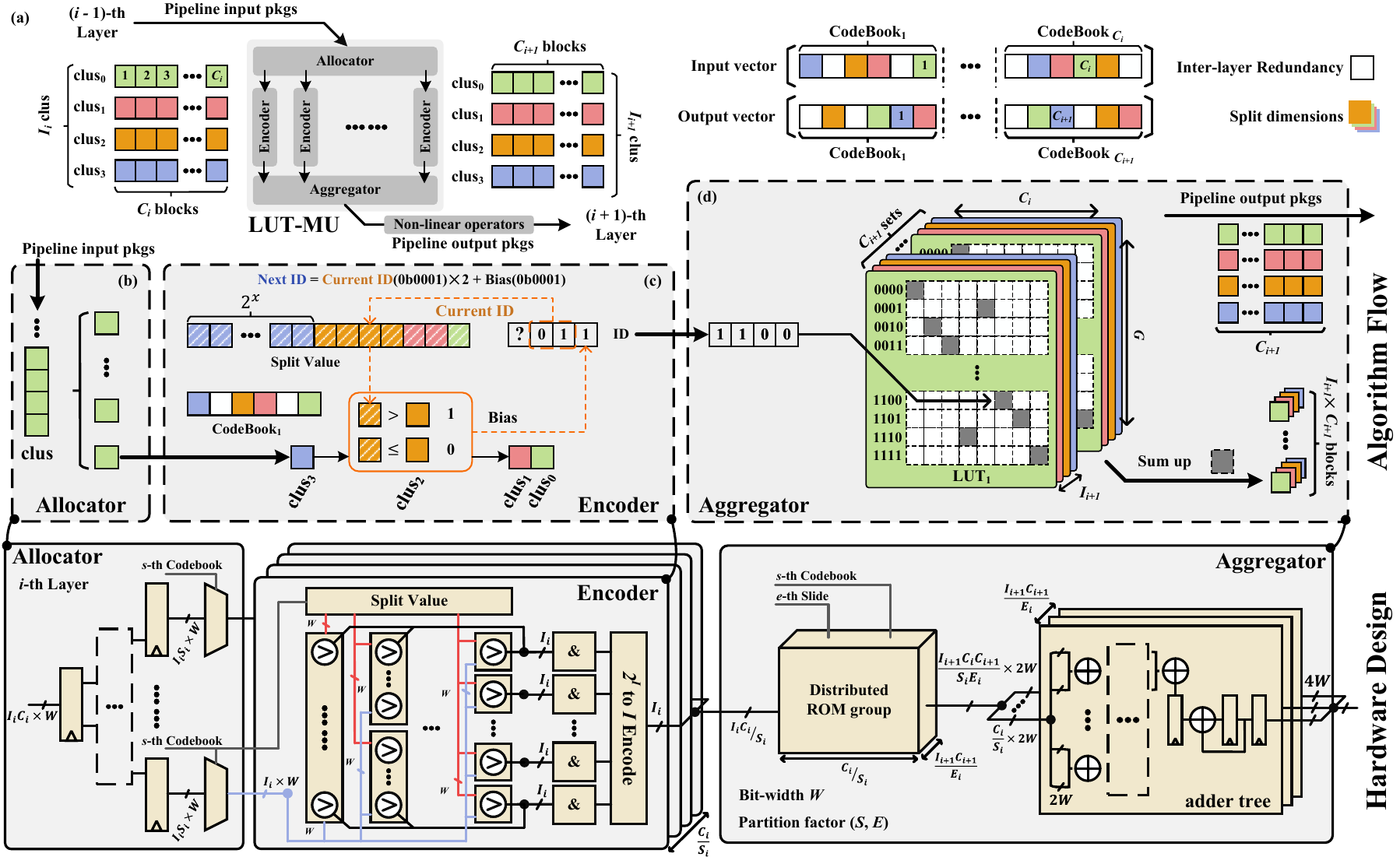}
    \caption{Overview of the proposed LUT-MU architecture: (a) The dataflow format and three components of LUT-MU: (b) Allocator; (c) Encoder; (d) Aggregator;}
    \label{fig:amu_overview}
\end{figure*}

However, directly eliminating redundancies in Im2col-based convolution is challenging due to two key issues. First, the $i$-th LUT is exclusively responsible for computing the $i$-th channel (i.e., the $i$-th element across all $L$ output points), but the split dimensions of the output feature map are randomly scattered across the $D_{out}$ channels. This irregularity makes channel-wise pruning infeasible. Second, when the convolution stride is smaller than the kernel width, the convolution windows overlap, causing the majority of cells in the output feature map to be marked as split dimensions. As a result, pruning becomes ineffective when most LUTs remain active throughout all $L$ rounds of multiplication. Therefore, an alternative approach is needed to transform the convolution operation into vector-matrix multiplication in a way that enables the use of MADDNESS properties for redundancy pruning.


As shown in Fig.~\ref{fig:kn2col}, to leverage the MADDNESS property, mitigating data movement overhead, and improving effective bandwidth, rather than flattening convolution windows into vectors (i.e. unfolded windows) via Im2col, it is more efficient to treat each convolution window as $K^2$ vectors (i.e., input points) along the $D_{in}$ channel dimension using Kn2col. In this manner, the split dimensions are gathered in some channels systematically, thereby enabling pruning optimisation. The samples in the training set $\tilde{A}$ can then be organised accordingly. By clustering the training set $\tilde{A}$, as described in Section~\ref{sec:offline}, the input points can share the same split dimensions during the inference phase. Consequently, redundant dimensions can be concentrated in specific channels of the input feature map, allowing easy pruning of these interlayer redundancies.
\vspace{-0.3cm}

\subsection{LUT-MU Architecture Overview}
LUT-MU is based on the MADDNESS architecture, consisting of an allocator, encoder, and aggregator. LUT-MU, however, optimises each computing block according to the three proposed optimisation techniques. These three processing units are responsible for allocating blocks, identifying prototypes, fetching partial results, and packing the multiplication results. In the following subsections, we first introduce the dataflow format of the LUT-MU, followed by a detailed discussion of the three processing units from both the algorithmic flow and hardware design perspectives.

\subsubsection{Dataflow format}

Fig.~\ref{fig:amu_overview}.(a) illustrates the dataflow format in algorithm flow, an LUT-MU in $i^{th}$ layer receives a pruned input vector with $I_i\times C_i$ elements from the previous layer via $I_i$ clusters, and sends $I_{i+1}$ clusters (each containing $C_{i+1}$ blocks) as a pruned output vector with $I_{i+1}\times C_{i+1}$ elements to subsequent non-linear operators before proceeding to the next layer. Here $C_i$ (or $C_{i+1}$) and $I_i$ (or $I_{i+1}$) denote that there are $C_i$ (or $C_{i+1}$) codebooks in the input (or output) vector and each codebook contains $I_i$ (or $I_{i+1}$) split dimensions. From the hardware design perspective, the input clusters of the $i$-th layer LUT-MU are assembled as a package, containing $I_i\times C_i$ elements with a $W$-bit data precision, where $W$ determines the quantisation level of activation or weight in inference networks. The resulting $I_{i+1}\times C_{i+1}$ elements are then delivered to the subsequent non-linear operator as $I_{i+1}\times C_{i+1}/E_i$ small package with a $2W$-bit data precision, where $E_i$ is a partition factor that influences bandwidth and resource usage (will be discussed in Section \ref{design_space}). The nonlinear operation then restores the results to $W$-bit data precision to ensure compatibility with subsequent computations.


\subsubsection{Allocator}

The main function of the allocator is to unpack the received package and allocate it to a series of parallel encoders. As shown in Fig.~\ref{fig:amu_overview}. (b), from the algorithm flow perspective, $I_{i}$ $C_i\times W$-bits-clusters (i.e. $I_{i}$ $clus$) will arrive successively due to data reshape in Section \ref{sec:opt_2} However, from a hardware design perspective, $I_i$ clusters can be assembled into a single data package to minimise throughput degradation caused by multiple package dispatches. Specifically, the $I_{i} \times C_i\times W$ data package is first fed into a tree-structured register group to generate a separate $I_i \times S_i\times W$ package. This tree-structured register group reduces fan-out, mitigating routing congestion and timing violations. The separated packages are then passed through a series of multiplexers (MUXs) to distribute the $C_i$ codebook encoding tasks among the $C_i/S_i$ encoders by $S_i$ rounds, where $S_i$ represents the partition factor (will be discussed in Section \ref{design_space}), which influences bandwidth and resource usage. 


\subsubsection{Encoder}\label{sec:encoder}


The $I_{i} \times C_i$ blocks are unpacked into $C_i/S_i$ corresponding encoders; each encoder is assigned $I_{i} \times S_i$ blocks to encode the IDs of the $S_i$ codebooks. These IDs are used to look up the dot product result between the corresponding prototype and the weight sub-matrix in the aggregator. As shown in the algorithm flow of Fig.~\ref{fig:amu_overview}(c), the encoder reads blocks serially to recursively process the decision tree-based encoding function. In each round, the encoder takes the current block from the input package along with the corresponding split value to determine the next round ID. Once all $I_i$ blocks have been processed, the encoder obtains an ID indicating the prototype address of the given codebook. However, from a hardware design perspective, data dependency in the recursive process leads to throughput degradation. To address this, the encoder uses $I_i$ comparator arrays to process all $2^{I_i}$ cases in the decision tree simultaneously, thus eliminating the data dependency. AND gates and a $2^{I_i}$-to-$I_i$ encode then identify the correct ID among the $2^{I_i}$ cases. This approach enables $I_i$ blocks to be loaded onto the encoder concurrently, enhancing the throughput of the LUT-MU.

\subsubsection{Aggregator}

In the algorithm flow of Fig.~\ref{fig:amu_overview}.(d), the case $C_i = 8$, $I_i = 4$ and $I_{i+1} = 4$ is demonstrated. The aggregator contains $I_{i+1}\times C_{i+1}$ LUTs obtained from the offline training stage, each LUT has $2^{I_i} \times C_i$ approximate partial product results. For each $CodeBook$, there is a distinct prototype ID ranging from $0b0000$ to $0b1111$ ($2^{I_i}$ prototypes in total). Considering that ID $1100$ is the prototype address of the $6$-th codebook (i.e. $CodeBook_{6}$), the cells in all $C_{i+1}\times I_{i+1}$ LUTs located in (12, 6) are selected as the partial dot product result of the input feature map and weight. All picked cells (marked as grey cells) are summed up, and bias is added if required. They are then passed through the following non-linear operator (i.e., successive thresholding operation in FINN framework \cite{Streamlined}, equivalent to scaling, batch normalisation, and uniform-quantised activation) to obtain $I_{i+1}\times C_{i+1}$ blocks and pipelined to send $I_{i+1}$ clusters to the next LUT-MU. From a hardware design perspective, the aggregator utilises a distributed ROM group and $I_{i+1}\times C_{i+1}/E_i$ parallel adder trees. The distributed ROM group, which stores the contents of $I_{i+1}\times C_{i+1}$ LUTs, consists of $(I_{i+1}\times C_{i+1} \times C_i)/(S_i\times E_i)$ ROMs. These ROMs are controlled by two selecting signals (i.e. s-th codebook and e-th slide, used to locate the correct column in the LUTs) and $(I_i \times C_i)/S_i$ address signals (i.e. IDs from Encoder). $(I_{i+1}\times C_{i+1} \times C_i)/(S_i\times E_i)$ partial product results are then simultaneously loaded onto $I_{i+1}\times C_{i+1}/E_i$ parallel adder trees from the distributed ROM group, where each adder tree includes an additional adder and a register group at the output stage to enable the summation of the $S_i$ rounds. To prevent overflow, the $2W$-bit precision partial results in LUTs are accumulated and extended by adder trees to produce the output of $4W$-bit precision.




\subsection{Hardware Design Space Exploration}\label{design_space}

\begin{figure}[h]
    \centering
    \includegraphics[width = \linewidth]{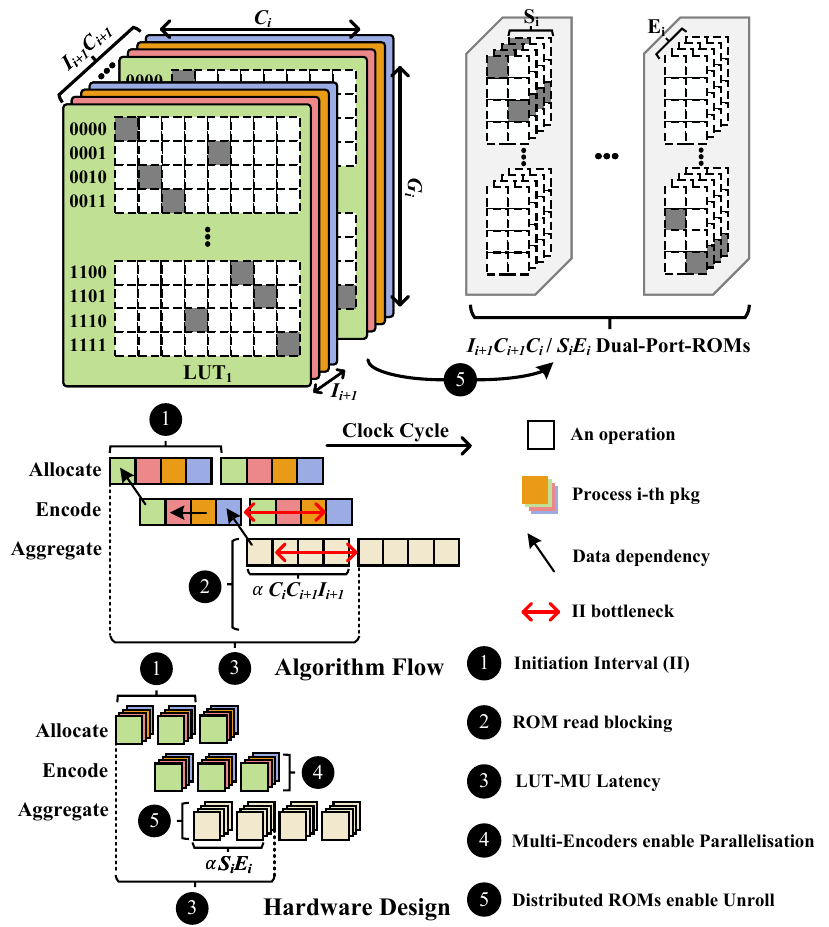}
    \caption{LUT-MU memory allocation and throughput (II) analysis.}
    \label{fig:hardware_design}
\end{figure}

Unlike GEMM, which uses element-by-element arithmetic operations, the LUT-MU uses incoherent memory accesses to achieve matrix multiplications. As shown in Fig.~\ref{fig:amu_overview} (c), LUT-MUs need to fetch the partial dot products (grey cells) scattered randomly across the LUT to obtain the approximated results. These random memory access behaviours hinder NNs from being computationally efficient: (1) Aggregating all necessary partial dot-product amounts from $C_{i}\times C_{i+1} \times 2^{I_i}$ can consume a large amount of on-chip resources. (2) The memory access contention of the encoding and aggregating processes can result in a pipeline stall. Therefore, the memory allocation of LUTs and the design of memory access strategies are crucial for achieving both the resource utilisation and throughput optimised LUT-MUs.


\subsubsection{Memory allocation design}\label{memory_allocation}

 As shown in Fig.~\ref{fig:amu_overview} (c), only $1/2^{I_i}$ gray cells in LUTs contribute to compute $I_{i+1}\times C_{i+1}$ approximate matrix multiplying results, while LUT-MU needs to pre-store $I_{i+1}\times C_{i+1}$ abstract LUTs. However, from a hardware design perspective, the random distribution of grey cells and limited read ports restricts rapid data access, thereby decreasing effective bandwidth and degrading throughput.
 To improve the effective bandwidth of the aggregator and enhance overall throughput, we analysed the structure of the LUT. As depicted in Fig.~\ref{fig:hardware_design}, each codebook can only be identified as a specific prototype, requiring only one partial dot product to be read from each column to compute a single element of the output (i.e., there is only one grey cell in each column of the LUT). Based on this feature, we partition the LUTs to various dual-port ROMs, each containing every $ S_i \times E_i$-columns partial dot product from LUTs, where $(S_i/2)| C_i $ and $E_i| (C_{i+1}\times I_{i+1}) $ are partition factors. 
 For example, a LUT-MU with $C_i$ input codebooks and $I_{i+1}\times C_{i+1}$ output elements can be assigned $(I_{i+1}\times C_{i+1}\times C_i) / (S_i \times E_i)$ dual-port ROMs. These dual-port ROMs, organised as a distributed ROM group, are driven by two selection signals $s$ and $e$ ($s < S_i$, $e<E_i$) to help locate the corresponding columns in LUTs for a given codebook ID (address signal). Adjusting $S_i$ and $E_i$ can reduce the scale of parallel encoders and adder trees, while optimising memory resource utilisation by minimising vacant regions in each ROM. More importantly, the distributed ROM group increases the number of read ports, thereby overcoming the bandwidth bottleneck caused by incoherent memory access patterns.




\subsubsection{Memory accesses pipeline design}\label{memory_accesses}




Although partitioning LUTs into distributed ROM groups enhances the theoretical bandwidth, the LUT-MU still requires a well-designed memory access architecture to boost effective bandwidth to its theoretical limit, thereby significantly improving throughput.
As shown in the algorithm flow of Fig.~\ref{fig:hardware_design}, the II of the LUT-MU is the time delay in cycles between the launch of processing successive input vectors. Two crucial bottlenecks prevent LUT-MUs from achieving lower II (i.e., the red double arrows in Fig.~\ref{fig:hardware_design}). Specifically, (1) sequentially allocating and encoding cluster results in $\alpha \times I_i$ clock cycle delay, and (2) aggregating partial dot product from $S_{i+1} \times I_{i+1}$ LUTs results in $\alpha \times S_i \times S_{i+1} \times I_{i+1} $ clock cycle delay. The final II of the LUT-MU is the maximum delay between two bottlenecks. 
To address the bottleneck of (1), which arises due to data dependencies both between cluster allocation and encoding, as well as among encoding clusters of an input vector, we can assemble all clusters and transmit them to $C_i/S_i$ parallel encoders by $S_i$ operations. This enables the encoder to process successive inputs in a pipeline. Additionally, to eliminate data dependencies caused by the recursive nature of the decision-tree encoding function, we utilise $2^{I_i}$ comparator arrays as described in Section \ref{sec:encoder}. The bottleneck of (2), caused by incoherent memory access patterns, can be addressed by using the distributed ROM group introduced in Section \ref{memory_accesses}. The distributed ROM group works in conjunction with parallel encoders and adder trees to avoid pipeline stalls and enhance effective bandwidth. 
However, there is $\alpha \times S_i \times E_i$ clock cycle latency in fetching all dot product results due to the read blocking of the ROM. Here, $\alpha$ denotes the average clock cycle delay of an operation in Fig.~\ref{fig:hardware_design}. To compress this delay, we can decrease $S_i$ or $E_i$. In an extreme scenario, we can allocate every column to distributed ROMs or separated registers so that all the grey cells can be fetched at once without read conflicts, which enables fully unrolling aggregating processes to achieve the best throughput.


\section{\deleted{Case study:} Implementation and Evaluation of LUT-MU \deleted{on FPGAs}}\label{section:result} 

To evaluate the performance of the proposed LUT-MU, we implement the LUT-MU module on AMD FPGA XCZU7EV and XCZU19EG devices with regard to the following four design perspectives: 1) assessing the effect of proposed pruning optimisations on NN inference performance; 2) evaluating the scalability of the single LUT-MU in the context of NN inference tasks with various LUT configuration and problem sizes; 3) exploring the optimal hardware setting for the best throughput and energy efficiency trade-off for a single LUT-MU module; 4) comparing the performance of the LUT-MU based NNs with the prior works. 
\vspace{-0.3cm}
\subsection{Deploying LUT-MUs-based QNNs on FPGA}

\begin{figure}[htb]
    \centering
    \includegraphics[width=\linewidth]{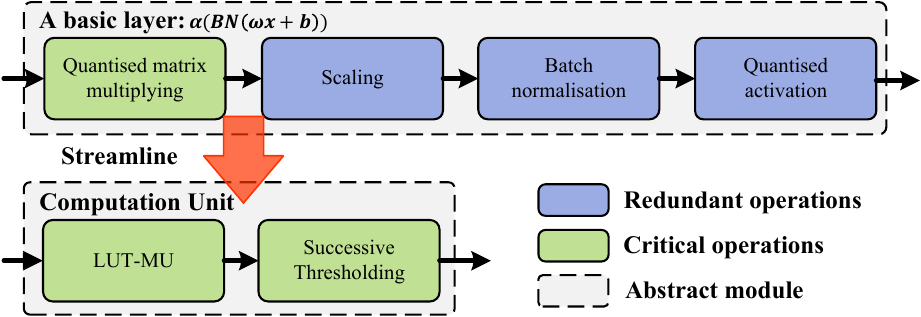}
    \caption{Using FINN pipeline \cite{Blott2018} to streamline basic layers for the LUT-MU based NNs.  }
    \label{fig:Deploying}
\end{figure}

To verify the LUT-MU performance, we employ the FINN pipeline as a development tool. FINN is an FPGA-oriented, streamlined QNN and end-to-end deployment tool. It can automatically generate customised FPGA-based computation modules for a quantised inference network. The key techniques of FINN can be summarised as three strategies: quantised-aware training, streamline, and folding strategies, which make FINN possible to transform an abstract inference network into a topological map cascaded by a series of reusable basic logic units for QNN computation: Sliding Windows Unit, Pool Unit, and MVAU. The topological map is then converted into a synchronous dataflow model by assembling basic computation units into corresponding independent modules and inserting dataflow-supported buffers. Finally, a QNN is mapped into a set of computation kernels on an FPGA.

As shown in Fig.~\ref{fig:Deploying}, we leverage the streamline strategy to eliminate redundant operations of the basic layer: scaling, batch normalisation, and activation operation, and substitute them with successive thresholding operations. Additionally, the LUT-MU is integrated into the computation unit as a substitution for quantised matrix multiplication by using offline training, as described in Section \ref{sec:offline}. These enable the LUT-MU-based computation unit to compute $\alpha(BN(wx+b))$ using only LUT-MU (i.e., approximate matrix multiplication) and successive thresholding operations. The convolution operation can be treated as a series of matrix multiplications, leveraging the img2col or kn2col function. Finally, an NN model can then be converted into cascaded LUT-MU-based computation kernels on an FPGA.

\vspace{-0.3cm}
\subsection{Evaluation of Algorithm Optimisation }

To assess the effect of proposed pruning optimisations on the NN inference performance (i.e., workload, parameter size, and inference accuracy), we utilise the 4-bit quantised ResNet-9 as the base model \footnote{Code is available in \url{https://github.com/EIS-Ressearch-Lab/LUT-MU}}. The quantised matrix multiplication units in 2nd to 7th layers of the base model are replaced with either the Halutmatmul (a LUT-based matrix multiplication unit derived from MADDNESS) units or LUT-MUs based units incorporating the proposed pruning optimisations under various LUT configurations $(d_{sub} \times 2^I)$, where $d_{sub}$ represents the codebook length and $2^{I}$ is the number of prototypes per codebook. Although the first convolution layer and the last fully connected layer remain 4-bit quantised exact matrix multiplication to mitigate accuracy loss. These ResNet-9 models, utilising LUT-based multiplication, are then layer-wise retrained and fine-tuned for 300 epochs by using Stella Nera \cite{halutmatmul} to recover the accuracy loss. However, the use of low-bit quantisation in the base model, combined with a limited number of codebooks and prototypes for approximating inputs, inevitably introduces noise. As a result, even after retraining, the LUT-MU-based model still exhibits a loss of accuracy of 4\% - 5\% compared to the base model (92.3\%).

\begin{figure}[htb]
    \centering
    \includegraphics[width=\linewidth]{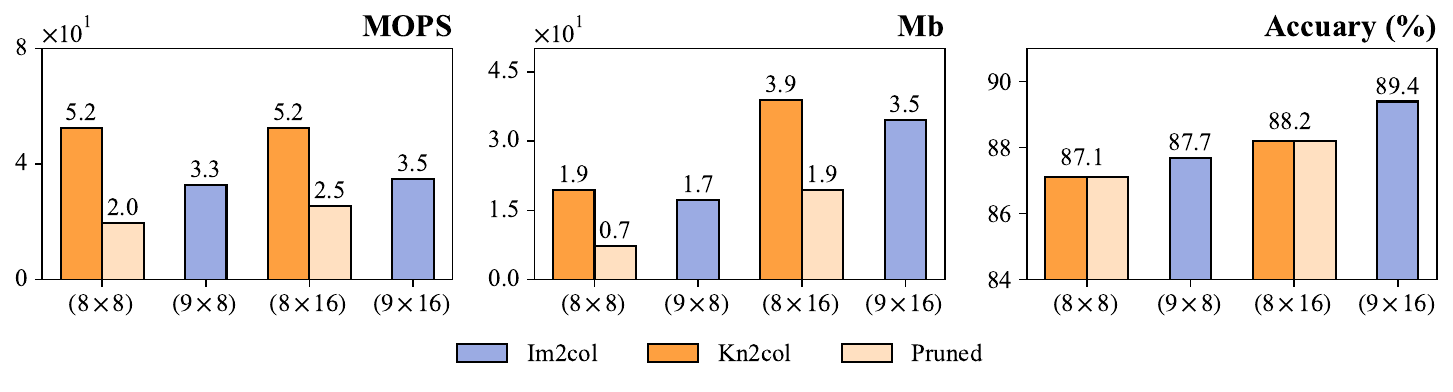}
    \caption{Original Halutmatmul (Im2col), Kn2col-based Halutmatmul (Kn2col), and LUT-MU with proposed pruning optimisations (pruned) implemented 4-Bit quantised ResNet-9 performance in various LUT configuration $(d_{sub} \times 2^{I})$ (i.e., codebook length $d_{sub}$ and number of prototypes per codebook $2^{I}$). \textbf{Left}: ResNet-9 workload in MOPs (Mega Operations), \textbf{Middle}: ResNet-9 parameter size in Mb (Mega bits), \textbf{Right}: Inference accuracy on CIFAR10.}
    \label{fig:param_acc_workload}
\end{figure}

The original Halutmatul (Im2col) has a default LUT configuration $d_{sub} = K\times K$ (kernel width $K = 3$ in ResNet-9) and $2^I$ prototypes per codebooks by using Im2col to unfold convolution windows, as shown in Fig. \ref{fig:kn2col}. Although Kn2col-based Halutmatmul (Kn2col) and LUT-MU (Pruned) have a $d_{sub} = D_{in}/C_{in}$ such that $D_{in}$ (number of the input channel) can be divided by $C_{in}$ (number of the input codebook). Therefore, the $d_{sub}$ of Kn2col-based Halutmatmul cannot be consistent with $d_{sub}$ of the original Halutmatmul. Here we use $d_{sub} = 8$ for LUT-MU and Kn2col-based Halutmatmul and $d_{sub} = 9$ for the original Halutmatmul, along with 8, 16, prototypes per codebook, and totally 4 different LUT configurations as examples of comparison.

Fig.~\ref{fig:param_acc_workload} illustrates the performance, parameter size, and accuracy comparisons. With similar configuration settings between Im2col $(9\times16)$ and Kn2col $(8\times16)$, LUT-MU with pruning optimisation reduces up to 58\% parameter size, compared to Im2col with minor accuracy loss (0.6 - 1.2\% loss).
The Kn2col-based Halutmatmul (Kn2col) using LUT configuration $(8\times16)$ and $(8\times8)$ has about 11\% parameter size expansion and 48\% - 57\% workload growth, compared to the original Halutmatmul (Im2col) using LUT configuration $(9 \times 16)$ and $(9 \times 8)$, respectively. Additionally, the Kn2col has 1.2\% accuracy loss compared to Im2col because of convolution layout differences between Kn2col and Im2col, as reported in \cite{halutmatmul}. The parameter size expansion, together with accuracy loss and workload growth, makes Kn2col less effective for NN inferences. In contrast, the LUT-MU with proposed pruning optimisation (pruned), using LUT configuration $(8\times 16)$ and $(8\times 8)$, achieves 46\% - 59\% parameter size compared to original Halutmatmul using LUT configuration $(9 \times 16)$ and $(9 \times 8)$, respectively, while simultaneously reducing workload without incurring additional accuracy loss unlike Kn2col. 

The analysis of Fig. \ref{fig:param_acc_workload} demonstrates that the proposed pruning optimisation effectively eliminates inter- and intra-layer redundancies in sequential LUT-based matrix multiplication computation units, thereby reducing the parameter size and computation workload without compromising accuracy. These improvements facilitate the development of more efficient hardware designs for LUT-MU.

\subsection{\added{Accuracy degradation investigation}}

\added{Understanding the impact of employing LUT-based approximate matrix multiplication (i.e., PQ) on neural-network accuracy is essential, as it guides designers towards minimising accuracy degradation while maximising performance gains. Tang et al. \cite{Tang2023} observed that approximation errors accumulate as more layers are replaced with LUT-based approximated matrix multiplications, ultimately leading to reduced accuracy. Moreover, the LUT configuration in each layer directly affects model accuracy. Therefore, we define resolution as the ratio between the number of split dimensions ($I$) and the codebook length ($d_{sub}$), denoted by $I/d_{sub}$. This resolution serves as an indicator of inference accuracy for a given LUT configuration. }

\begin{figure}[htbp]
    \centering
    \includegraphics[width=\linewidth]{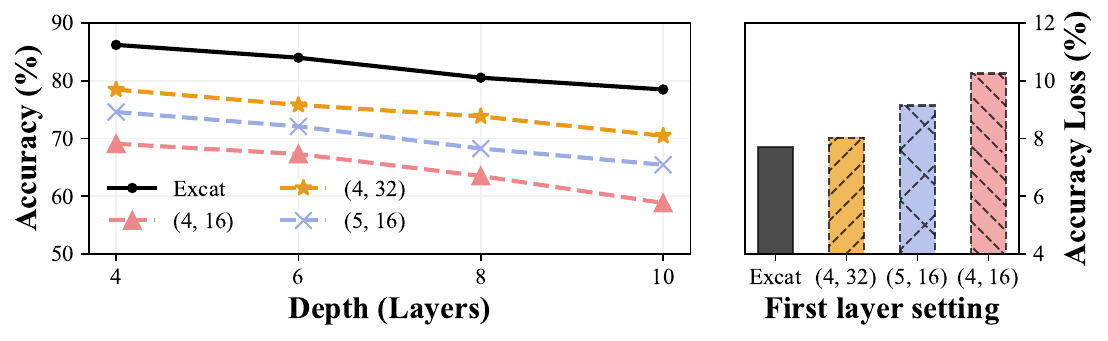}
    \caption{\added{The impact of prefix layer configurations on LUT-MU-based MLP accuracy as depth increases, using the first layer setting as an example. 
    ``Exact'' is `Exact matrix multiplication', the pair $(C, ~2^I)$ represent $C$ codebooks and $2^I$ prototypes for the first layer (The rest layers with setting (4, 16))}}
    \label{fig:accuracy_decay}
\end{figure}

\added{Intuitively, the upper bound of model accuracy is constrained by the layer with the lowest resolution. However, as shown in Fig.~\ref{fig:accuracy_decay}, early layers are also critical: in a 4-layer fully connected MLP, using a lower-resolution configuration in the first layer results in a lower accuracy baseline (e.g., MNIST classification accuracy) and causes accuracy to degrade more rapidly as the depth increases. Specifically, we employ MLPs that contain a $784 \times 256 $ matrix multiplication as the first layer, followed by repeated hidden layers (a hidden layer includes: $256\times 256$ matrix multiplication, batch normalisation, activation, and dropout) and $256 \times 10$ classification layer. These MLPs are then implemented by LUT-MU with the same resolution (i.e., 4 codebooks, 16 prototypes per codebook) for all hidden layers and the classification layer, except the first layer with 4 different configurations: (4, 16), (4, 32), (5, 16) and Exact matrix multiplication (i.e., 32-bit floating point arithmetic). Where the Exact matrix multiplication can be considered as the upper boundary of the LUT-MU when $N = 28 \times 28$, $I \to \infty$. In principle, the length of codebooks ($d_{sub}$) is allowed to be unequal if the number of channels cannot be decided by $C$. However, in practice, unequal $d_{sub}$ makes hardware design more complex. In summary, for the NN using LUT-based approximated matrix multiplication, using higher resolution in the early layer can provide better accuracy and a better effect on alleviating accuracy loss with increasing depth. Similarly, to further raise the upper bound of accuracy, increasing the resolution in the subsequent layer is required.}


\subsection{Scalability of the LUT-MU}

\begin{figure}[h]
    \centering
    \includegraphics[width = \linewidth]{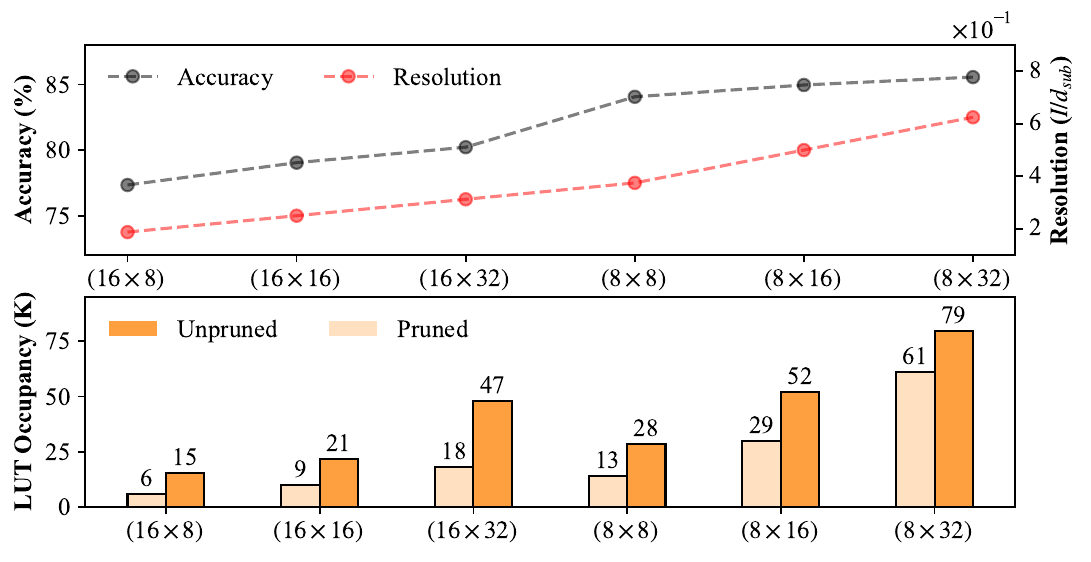}
    \caption{Pruned/Unpruned LUT-MU scalability as the accuracy grows. \textbf{Top}: 25 epochs \deleted{fine-tuning} \added{layer-wise retrained} accuracy of LUT-MU based 4-bit quantised ResNet-9 on CIFAR10 across varying resolutions ($I/d_{sub}$); \textbf{Bottom}: Layer 2 LUT-MU LUT (FPGA resource) utilisation profiling for different LUT configuration ($d_{sub}\times2^I$).}
    \label{fig:occupancy}
\end{figure}

In this section, we explore the scalability of the proposed design from two perspectives: the growth of resource usage with increasing inference accuracy and with increasing problem size. To evaluate the impact of accuracy on resource consumption, \deleted{we measure the LUT occupancy (FPGA resource) of the second layer LUT-MU in a 4-bit quantised ResNet-9 model, where the LUT-MU based ResNet-9 model is retrained for 25 epochs using six different LUT configurations $(d_{sub} \times 2^I)$.}
\added{we measure the LUT occupancy (FPGA resource) of the second layer LUT-MU in a 4-bit quantised ResNet-9 model, where the LUT-MU based ResNet-9 models substitute all exact matrix multiplications, except those in the first and the last layer, with LUT-MUs using six different LUT configurations $(d_{sub} \times 2^I)$, and are layer-wise retrained for 25 epochs as described in \cite{halutmatmul}. }
\deleted{We define resolution as the ratio between the number of split dimensions ($I$) and the codebook length ($d_{sub}$), denoted by $I/d_{sub}$.  This resolution serves as an indicator of inference accuracy for a given LUT configuration. }
As shown on the top of Fig.~\ref{fig:occupancy}, the inference accuracy of the LUT-based model has a positive correlation with the resolution. This is because, the accuracy loss of the LUT-MU base model with sufficient retraining depends mainly on the LUT configuration (i.e., $I/d_{sub}$). 

As illustrated in Fig.~\ref{fig:occupancy}, our pruning shows $1.3 \text{-} 2.6\times$ LUT usage efficiency, depending on variable configurations. Additionally, it reduces the growth in LUT occupancy by 5\% to 63\% as accuracy requirements (resolution) increase, for configurations with $d_{sub} = 8$ and $d_{sub} = 16$, respectively.
In terms of accuracy, there is a significant boost when moving from a $(16 \times 32)$ configuration to $(8 \times 8)$, indicating that the sub-vector dimension is the primary factor influencing accuracy when the number of prototypes ranges from 8 to 32.
LUT occupancy fluctuates between 6 K and 61 K for pruned LUT-MU, and between 15 K and 75 K for unpruned LUT-MU, due to changes in parameter size. Specifically, the LUT occupancy of the pruned LUT-MU increases by 12 K compared to 32 K for the unpruned version when scaling from $(16 \times 8)$ to $(16 \times 32)$, and by 48 K compared to 51 K for the unpruned version when scaling from $(8 \times 8)$ to $(8 \times 32)$. However, the pruned LUT-MU exhibits a 51 K increase from $(8 \times 8)$ to $(8 \times 32)$, which is significantly greater than the 12 K growth in the $d_{sub} = 16$. This indicates that the LUT configuration affects the pruning efficiency, and increasing $d_{sub}$ can reduce LUT occupancy growth rate as resolution increases. However, the accuracy comparison between the LUT configuration with $d_{sub} = 8$ and $d_{sub} = 16$ reveals that increasing $d_{sub}$ degrades approximation precision with a fixed number of prototypes, thus reducing inference accuracy. Theoretically, the LUT-MU with $C/2$ codebooks (i.e., $2\times d_{sub}$ codebook length) needs $2^{(2\times I)}$ prototypes (i.e., $2\times I$ split dimensions) to restore accuracy as the LUT-MU with $(d_{sub} \times 2^I)$ LUTs configuration. However, resource-constrained devices are incapable of tolerating exponential growth in LUTs size. Therefore, we need to provide an appropriate number of prototypes to meet the resource constraints of those devices.

\begin{figure}[h]
    \centering
    \includegraphics[width = \linewidth]{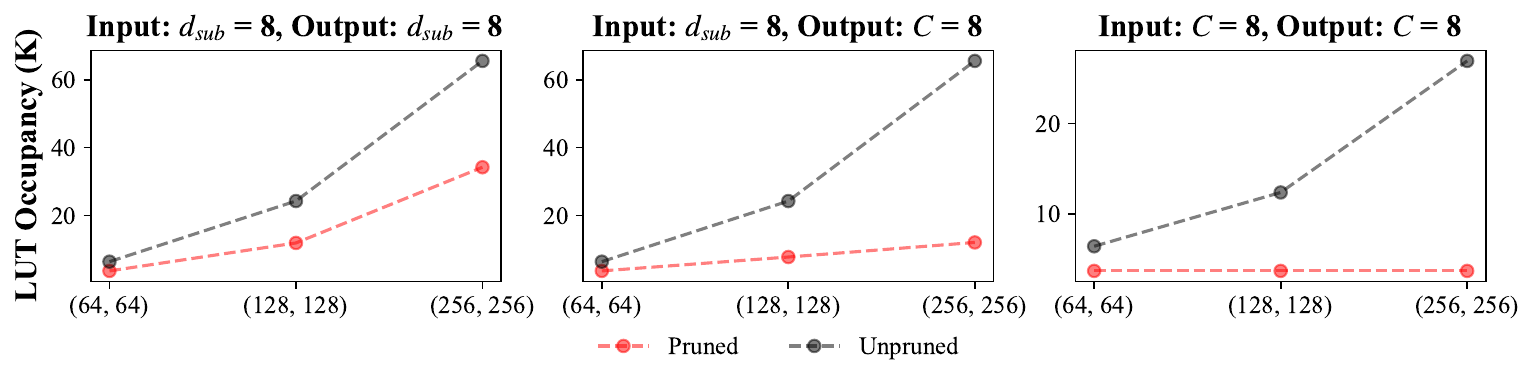}
    \caption{The impact of problem sizes $(D_{in},D_{out})$ on the LUT resource occupancy for a single LUT-MU with three different LUT configurations on the XCZU19EG platform@100 MHz. \textbf{Left}: Both input dimension $D_{in}$ and output dimension $D_{out}$  uses fixed codebook length $d_{sub} $; \textbf{Middle}: input dimension $D_{in}$ uses fixed codebook length $d_{sub}$, output dimension $D_{out}$ uses fixed number of codebook $C$; \textbf{Right}: Both input dimension $D_{in}$ and output dimension $D_{out}$ uses fixed number of codebooks $C$. All LUT configurations use $2^I = 16$ prototypes in each codebook.}
    \label{fig:problem_size}
\end{figure}

From another perspective, pruning optimisation significantly improves the resource scalability of the LUT-MU in increasing the matrix multiplication problem size. As shown in Fig. \ref{fig:problem_size}, we investigated the impact of varying problem sizes on the LUT occupancy of a single LUT-MU, using three different LUT configurations. When the size of the problem increases from $(64, 64)$ to $(256, 256)$, the LUT occupancy of the unpruned LUT-MU increases by 33K, while the pruned LUT-MU only increases by 18K, if both the input and output dimensions use $d_{sub} = 8$. Furthermore, comparing the pruned LUT-MU with output $d_{sub} = 8$  against that with output $C = 8$ reveals that fixing the number of codebooks in the output dimension significantly mitigates LUT resource growth as the problem size. When a fixed number of codebooks $C$ is used for both input and output dimensions, the unpruned LUT-MU still suffers from increased resource usage due to growing parameter redundancy. In contrast, the LUT occupancy of pruned LUT-MU remains nearly constant. This is because LUT-MU with pruning optimisation always needs to process the identical input information volume and send the same output information volume once $C$ and $2^I$ are decided. However, using the same hyperparameters for the larger problem size inevitably results in a loss of accuracy because of insufficient codebooks for describing the more sophisticated feature distribution.

The analysis of Fig.~\ref{fig:occupancy} and Fig.~\ref{fig:problem_size} reveals that MADDNESS and its derivative LUT-based matrix multiplication expand rapidly in resource costs along with the higher accuracy requirement (i.e., resolution) and larger problem size (i.e., the weight matrix size). In contrast, the LUT-MU with pruning optimisation eliminates the inter and intra-layer redundancies of LUT-based matrix multiplying, which offers a significant improvement on the resource scalability while maintaining consistent accuracy as the original MADDNESS, along with increasing the accuracy requirement and problem sizes. However, constraining resource growth comes at a cost: keeping the number of codebooks $C$ and prototype $2^I$ fixed while increasing $D_{in}$ and $D_{out}$ reduces resolution, which in turn degrades inference accuracy for neural networks with larger matrix multiplication problem sizes.

\subsection{Evaluation of Hardware Design Efficiency}

\begin{table}[htbp]\renewcommand\arraystretch{1.2}
  \centering
  \begin{threeparttable}
  \caption{LUT-MU based V.S MVAU based 1-bit quantised SFC model implementations on XCZU7EV@100 MHz.}
  \setlength{\tabcolsep}{8pt}{
    \begin{tabular}{cccc}
    \toprule
    \textbf{Computation Unit} & \multicolumn{2}{c}{\textbf{LUT-MU}} & \textbf{MVAU} \\
    \midrule
    \textbf{Accuracy} & 90.2\%  & 92.3\%  & 97.8\% \\
    \textbf{LUT}/\textbf{Weight Shape}\tnote{1} & (32, 8, 48) & (32, 16, 64) & (256, 256) \\
    \bottomrule
    \end{tabular}%
    }\label{tab:memory_design}
    \begin{tablenotes}
        \footnotesize
            \item[1] LUT shape for LUT-MU: $(C_{in}, 2^{I_{in}}, C_{out}\times{I_{out}})$;  Weight shape for MVAU: ($D_{in}$, $D_{out}$).
    \end{tablenotes}
    \end{threeparttable}
\end{table}%

\begin{figure}[h]
    \centering
    \includegraphics[scale = 0.5]{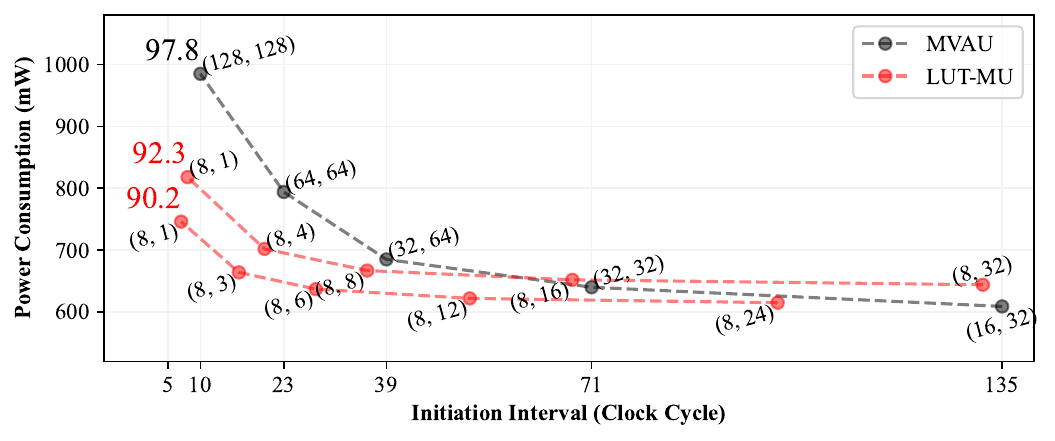}
    \caption{Pareto Optimal of the LUT-MU and MVAU on XCZU7EV @ 100 MHz with 5 different partition configurations. LUT-MU partition configuration $(S, E)$, MVAU partition configuration $(PE, SIMD)$}
    \label{fig:patero}
\end{figure}

To explore the memory allocation and access design for achieving the best throughput and energy efficiency of the single LUT-MU module, we adopt the SFC model, a four-layer fully connected neural network used for handwritten digit classification, as the target model. Specifically, we evaluate the performance of LUT-MU and MVAU implementations on the second layer. All modules are tested on the same problem size (i.e., second layer of SFC model with (256, 256) weight matrix or 0.13 MOPs). We utilise the Xilinx Power Estimator Tool to estimate the power consumption by giving the corresponding resource utilisation and clock frequency. As shown in Table~\ref{tab:memory_design}, the LUT-MU is implemented using two LUT shape configurations, denoted $(C_{in}, 2^{I_{in}}, C_{out} \times I_{out})$. These LUT-MU configurations form Pareto-optimal curves characterising the trade-off between power consumption and II, and are compared against MVAU implementations. Both LUT-MUs and MVAU are configured with 5 different partition factors: $(S, E)$ or $(PE, SIMD)$. The $(S,E)$ indicates the partition factor of LUT-MU such that $(S/2)$ divides $C_{in}$ and $E$ divides $(C_{out} \times I_{out})$. Specifically, the LUTs can be allocated every $S\times E$ columns as a block to each Dual-port-ROM, as discussed in Section \ref{design_space}. Additionally, 5 different MVAUs with varying partition factor $(PE, SIMD)$ are quoted as a baseline. Here, $PE$ and $SIMD$ denote the parallelism level of the MVAU, which influences the theoretical throughput (i.e. FPS) of the MVAU based on Eq.~(\ref{math:fps}), where $F_{clk}$ is the device clock frequency, and $F$ is the total fold defined in Eq.~(\ref{math:Fold}) \cite{Blott2018}.
\begin{align}
    & FPS = F_{clk}/F = F_{clk}/II \label{math:fps}\\
    & F = D_{in}/SIMD \cdot D_{out}/PE \label{math:Fold}
\end{align}

\begin{table*}[h]\renewcommand\arraystretch{1.2}
  \centering
  \begin{threeparttable}
  \caption{\added{Comparison with previous works}}
  \setlength{\tabcolsep}{4.5pt}{
  \begin{tabular}{ccccccccccc}
    \toprule
    \multirow{2}[2]{*}{} & \multirow{2}[1]{*}{\textbf{Platform}} & \multirow{2}[1]{*}{\textbf{Dataset}} & \multirow{2}[1]{*}{\textbf{Precision}} & \multirow{2}[1]{*}{\textbf{Technique}\tnote{1}} & \textbf{Model} & \textbf{Accuracy} & \textbf{Complexity} & \textbf{Frequency} & \textbf{Throughput} & \textbf{Efficiency} \\
          &       &       &       &       & \textbf{Structure} & \textbf{(\%)} & \textbf{(MOPs)} & \textbf{(MHz)} & \textbf{(GOPS)} & \textbf{(GOPS/W)} \\
    \midrule
    \multirow{3}[1]{*}{Our} & XCZU19EG & CIFAR10 & INT4  & QAT + PQ & ResNet-9 & 87.7  & 570   & 200   & 3,742 & 405 \\
          & XCZU19EG & IMAGENET & INT4  & QAT + PQ & ResNet-18 & 52.1  & 1,291 & 200   & 2,586 & 297 \\
          & XCZU19EG & IMAGENET\tnote{2} & INT4  & QAT + PQ & ResNet-50 & 60.0    & 2,518 & 200   & 2,442 & 305 \\
    \midrule
    \multirow{3}[1]{*}{\cite{Blott2018}} & XCZU19EG & CIFAR10 & INT4  & QAT + E2E & ResNet-9 & 92.5  & 570   & 200   & 3,490 & 231 \\
          & XCZU19EG & IMAGENET & INT4  & QAT + E2E & ResNet-18 & 54.3  & 1,291 & 200   & 2,261 & 238 \\
          & XCZU19EG & IMAGENET\tnote{2} & INT4  & QAT + E2E & ResNet-50 & 63.9  & 2,518 & 200   & 2,230 & 210 \\
    \midrule
    \multirow{3}[1]{*}{\cite{tensorrtwp}} & Jetson AXG Orin & CIFAR10 & FP16  & PTQ + E2E & ResNet-9 & 93.6  & 570   & 1,300  & 2,494 & 207 \\
          & Jetson AXG Orin & IMAGENET & FP16  & PTQ + E2E & ResNet-18 & 56.4  & 1,291 & 1,300  & 2,220 & 116 \\
          & Jetson AXG Orin & IMAGENET\tnote{2} & FP16  & PTQ + E2E & ResNet-50 & 65.6  & 2,518 & 1,300  & 1,559 & 73 \\
    \midrule
    \cite{PQA} & Agilex DE10 & CIFAR10 & FP16  & PTQ + PQ & ResNet-20 & 84.4  & 81    & 490   & 2,317 &  -  \\
    \cite{fuketa}\tnote{3} & XCZU7EV & IMAGENET & INT8  & PTQ + PQ & ResNet-18 & 64.5  & 310   & 150   & 620   & 88 \\
    \midrule
    \cite{angleye} & XC7Z045 & IMAGENET & INT16 & PTQ + E2E & VGG-16 & 67.8  & 552   & 150   & 137   & 14 \\
    \cite{dnnbuilder} & ZC7Z045 & IMAGENET & INT8  & QAT + NAS & AlexNet & 53.4  & 1,450 & 200   & 494   & 68 \\
    \bottomrule
    \end{tabular}%
}\label{tab:compare}
    \begin{tablenotes}[para]
        \footnotesize
            \item[1] QAT: quantisation‑Aware Training; PTQ: Post-training quantisation; E2E: End-to-End optimisation; NAS: Network Architecture Searching.
            \item[2] Tiny IMAGENET with 100 categories classification. 
            \item[3] Layer 3 performance only, including data transfer and preparation time.
    \end{tablenotes}\vspace{-0.5cm}
\end{threeparttable}
\end{table*}%

As shown in Fig.~\ref{fig:patero}, the LUT-MU with dedicated memory allocation and access design pushes the power-throughput optimal curve further forward by trading off acceptable accuracy loss and thereby boosting NN inference efficiency significantly compared to the quantised matrix multiplication unit MVAU. Specifically, the LUT-MU with 92.3\% accuracy has a similar II as the MVAU with 97.8\% in different partition configurations; however, the LUT-MU has 90 - 240 mW lower power consumption compared to MVAU when their II is lower than 40 clock cycles. Additionally, the LUT-MU with (S = 8, E = 1) achieves II = 8, while the MVAU with (PE = 128, SIMD = 128) fails to break through the practical boundary II = 10 to reach the theoretical II = 4 at $F=4$. This observation indicates that by trading off accuracy, the LUT-MU can achieve better energy efficiency when the throughput requirements become strict. From another perspective, the LUT-MU with (32, 8, 48) LUT shape (parameter volume $32\times8\times48 =  12,288$) trade-off 2\% accuracy with 37\% parameter volume compared to the LUT-MU with (32, 16, 64) LUT shape (parameter volume $32\times 16\times 48 = 32,768$) pushing Pareto optimal curves further forward, This observation suggests that reducing parameter volume can effectively improve the performance of LUT-based matrix multiplication unit in terms of resource occupancy, power consumption and throughput.

The analysis of Table \ref{tab:memory_design} and Fig.~\ref{fig:patero}, suggests that eliminating the inter and intra-layer redundancies enables the LUT-MU to obtain significant improvements in energy efficiency and throughput, compared to the MVAU \cite{Umuroglu2017}, by aiding with suitable memory allocation and memory access design for the effective bandwidth degradation derived from incoherence memory access behaviours introduced in Section \ref{sec:challenges}.


\subsection{Performance comparison}

\begin{table}[htbp]\renewcommand\arraystretch{1.2}
  \centering
  \begin{threeparttable}
  \caption{\replaced{The resource utilisation of LUT-MU based NN models}{The basic information of LUT-MU based NN Models}}
  \setlength{\tabcolsep}{8pt}{
    \begin{tabular}{ccccc}
    \toprule
    \textbf{Model} & \textbf{Complexity} & \multicolumn{3}{c}{\textbf{Resource Utilisation}} \\
    \textbf{Structure} & \textbf{(MOPs)} & \textbf{LUT} & \textbf{BRAM} & \textbf{FF} \\
    \midrule
        ResNet-9 & 570   & 204,569 & 518 @ 18k & 65,906 \\
        ResNet-18 & 1,291 & 167,409 & 610 @ 18k & 87,116 \\
        ResNet-50 & 2,518 & 148,637 & 840 @ 18k & 65,917 \\
    \bottomrule
    \end{tabular}}\label{tab:basic_info}
  \end{threeparttable}
\end{table}%

\deleted{We implement LUT-MU based NNs, including the SFC (a four-layer MLP), ResNet-9, and ResNet-18, on XCZU7EV and XCZU19EG Xilinx devices for evaluation. To make LUT-MU compatible with convolution computation, the convolution operations are transformed into a series of matrix multiplications using the kn2col. Specifically, 100\%, 99\%, and 37\% of the quantised matrix multiplication operations in the original 4-bit quantised models (MVAU-based) were replaced with LUT-MU modules in SFC, ResNet-9, and ResNet-18, respectively. As illustrated in Table \ref{tab:basic_info}, the LUT-MU-based model reduces operations by more than 95\% compared to the original 4-bit quantised model (MVAU-based) using quantised matrix multiplication. However, the LUT-MU-based model exhibits about 5\% accuracy loss compared to the MVAU-based baseline.}

\added{We implement LUT-MU based NNs, including the ResNet-9, ResNet-18, and ResNet-50 on XCZU19EG AMD Xilinx devices for evaluation. In order to make LUT-MU compatible with convolution computation, the convolution operations are transformed into a series of matrix multiplications using the kn2col. The resource utilisation of matrix multiplications implemented by LUT-MU on  XCZU19EG is illustrated in Table \ref{tab:basic_info}. Specifically, 99\% (570 MOPs), 37\% (1291 MOPs) and 31\% (2518 MOPs) of the quantised matrix multiplication operations in the original 4-bit quantised models (MVAU-based) were replaced with LUT-MU modules in ResNet-9, ResNet-18, and ResNet-50, respectively. The original models and MVAU-based quantised models deployed by tensorRT \cite{tensorrtwp} and FINN \cite{Blott2018} are trained from scratch for 300 epochs using a learning rate of $1\times 10^{-3}$ and a batch size of 256. The LUT-MU based NN models are derived from trained quantised models using the retrain strategy proposed by Jannis et al. \cite{halutmatmul}: each matrix-multiplication layer is replaced with a LUT-based approximate matrix-multiplication and 25 epochs layer-wise retrained, followed by a 300 epochs final fine-tuning.}

As illustrated in Table \ref{tab:compare}, the proposed work, together with \deleted{8 previous work} \added{6 previous works} implemented on different platforms with \added{different optimisation techniques and} varying frequencies, is validated by different task complexities (i.e., the total computing operations of the neural network). Peak throughput and peak efficiency are recorded as key criteria for demonstrating the performance of each NN implementation in this table, where throughput uses equivalent computation complexity to calculate throughput for LUT-based multiplication. 


\deleted{By trading about a 5\% accuracy drop compared to the MVAU-based design, the LUT-MU based design on XCZU19EG outperforms MVAU-based \cite{Blott2018} NN accelerators under the same hardware platform and computational complexity (SFC, ResNet-9, and ResNet-18), achieving $1.1\times$ to $2\times$ higher throughput and $1.8\times$ to $10\times$ greater energy efficiency. Furthermore, it reduces LUT occupancy by 20\% to 43\% compared to the MVAU-based design while maintaining comparable BRAM usage. From the other perspective, although the target models differ, the LUT-MU based design achieves a $3\times $ and $27 \times$ gain in throughput and a $3.5\times$ and $28 \times$ improvement in energy efficiency, compared to the previous works \cite{dnnbuilder} and \cite{angleye}, respectively, under comparable model complexities (552 MOPs vs. 570 MOPs and 1284 MOPs vs. 1450 MOPs). Furthermore, the proposed design achieves $2.6\times$ and $2.8\times$ improvements in throughput and energy efficiency, respectively, compared to \cite{fuketa}, a MADDNESS-derived matrix multiplication design based on LUT. These results indicate that the proposed pruning optimisations, combined with the hardware design, effectively enhance both throughput and energy efficiency.}

\added{By trading about a 5\% accuracy drop compared to the MVAU-based design, the LUT-MU based design outperforms MVAU-based \cite{Blott2018} NN accelerators under the same hardware platform and computational complexity (ResNet-9, ResNet-18, and ResNet-50), achieving a $1.3\times$ to $1.8\times$ improvement in energy efficiency, while maintaining comparable throughput. Furthermore, the LUT-MU-based design outperforms CUDA-based acceleration on Jetson AGX Orin 64GB under the same computation complexity, delivering $1.2\times$ to $1.6\times$ higher throughput and $1.9\times$ to $4.2\times$ greater energy efficiency by trading about a 6\% accuracy loss compared to the FP16 models deployed by tensorRT \cite{tensorrtwp}. }


\added{From the other perspective, LUT-MU can be applied to models with larger computation complexity, ranging from (570 MOPs to 2518 MOPs), while still achieving higher throughput compared to PQA \cite{PQA}. Additionally, the proposed design achieve $2.6\times$ and $2.8\times$ improvements in throughput and energy efficiency, compared to RLDA \cite{fuketa}, a LUT-based approximated matrix multiplication design derived from MADDNESS. RDLA, which employs a dedicated retraining strategy and uses additional LUTs to store residual product quantisation, achieves notable accuracy improvements, while it incurs significant expansion in memory footprint, resulting in significance throughput and efficiency degradation. Furthermore, although the target models differ, the LUT-MU based design achieves a $3\times $ and $27 \times$ gain in throughput and a $3.5\times$ and $28 \times$ improvement in energy efficiency, compared to the previous works \cite{dnnbuilder} and \cite{angleye}, respectively, under comparable model complexities (552 MOPs vs. 570 MOPs and 1284 MOPs vs. 1450 MOPs). }

However, when the complexity of the model computation increases, the throughput and efficiency of LUT-MUs for \deleted{ResNet-9 and ResNet-18} \added{ResNet-18 and ResNet-50} exhibit a significant drop compared to the \deleted{SFC} \added{ResNet-9}. This phenomenon can be attributed to two main reasons: 1) The increase in computation complexity is primarily derived from the expansion of feature map width, whereas the proposed pruning optimisations are only effective in mitigating performance degradation caused by increasing problem size along the codebook dimension. 2) To alleviate accuracy loss and increase resource demands, a \deleted{larger number of codebooks ($C$)} \added{higher resolution} and \added{larger} partition factors ($E$, $S$) are required for each layer. These observations indicate that the proposed LUT-MU still faces two main challenges: 1) the proposed pruning optimisations are only effective in minimising the performance degradation that results from increasing problem size along the dimension of the codebook. 2) The need for \deleted{additional codebooks $C$} \added{higher resolution} and larger partition factors ($E$, $S$) to mitigate accuracy loss and address increasing resource demands offset the gains in throughput and energy efficiency.


\section{Conclusion}\label{section:conclusion} 

\deleted{In this paper, we propose a novel LUT-based approximate matrix multiplication unit: LUT-MU, a scalable and energy-efficient LUT-based approximate matrix multiplication unit to mitigate the scalability issues inherent from LUT-based neural networks by pruning optimisations. By incorporating pruning optimisation and hardware-software co-design, the proposed LUT-MU-based NNs achieved up to $2\times$ higher throughput and up to $10\times$ greater energy efficiency than FINN-generated MVAU-based NN implementations that use bit-wise and pop-count operations on the quantised elements for the NN inference task with the same computation complexity and hardware platform, while sacrificing minor accuracy loss. Our future work involves more effective training methods and pipeline architectures for LUT-MU to minimise accuracy loss while managing bandwidth demands in complex NNs.}

\added{In this paper, we propose a novel LUT-based approximate matrix multiplication unit: LUT-MU, a scalable and energy-efficient LUT-based approximate matrix multiplication unit to mitigate the scalability issues inherent from LUT-based neural networks by pruning optimisations. By trading about 6\% accuracy loss, the proposed LUT-MU-based neural networks together with pruning optimisation achieved up to $1.6\times$ higher throughput and up to $4.2\times$ greater energy efficiency than TensorRT-optimised CUDA acceleration on the Jetson AGX Orin 64 GB at the same computation complexity. Under the same hardware platform and model complexity, the design delivers up to $1.8\times$ energy efficiency over FINN-generated MVAU-based quantised NN implementations while maintaining comparable throughput. However, the accuracy degradation limits its applicability to high-accuracy tasks, where the accuracy loss is mainly due to the inherent attribute of LUT-based approximated matrix multiplication, accumulated error along with increasing model computation complexity, and additional noise introduced by quantisation-aware training. Therefore, our future work involves exploring effective training method including: optimised training hyper-parameter configurations LUT-compatible quantisation-aware training and clustering strategies to minimise accuracy loss in the LUT-MU based quantised model.}


\section*{Acknowledgment}
This work is supported by the UK Engineering and Physical Sciences Research Council through grants EP/X015955/1, EP/X019160/1, EP/V000462/1, EP/V034111/1 and EP/Z533749/1. For open access, the author has applied
a Creative Commons Attribution (CC BY) licence to any Author Accepted
Manuscript version arising.


%



\ifCLASSOPTIONcaptionsoff
  \newpage
\fi


\bibliographystyle{IEEEtran}
\bibliography{ref}


\begin{IEEEbiography}[{
\includegraphics[width=1.0in,height=1.5in,clip,keepaspectratio]{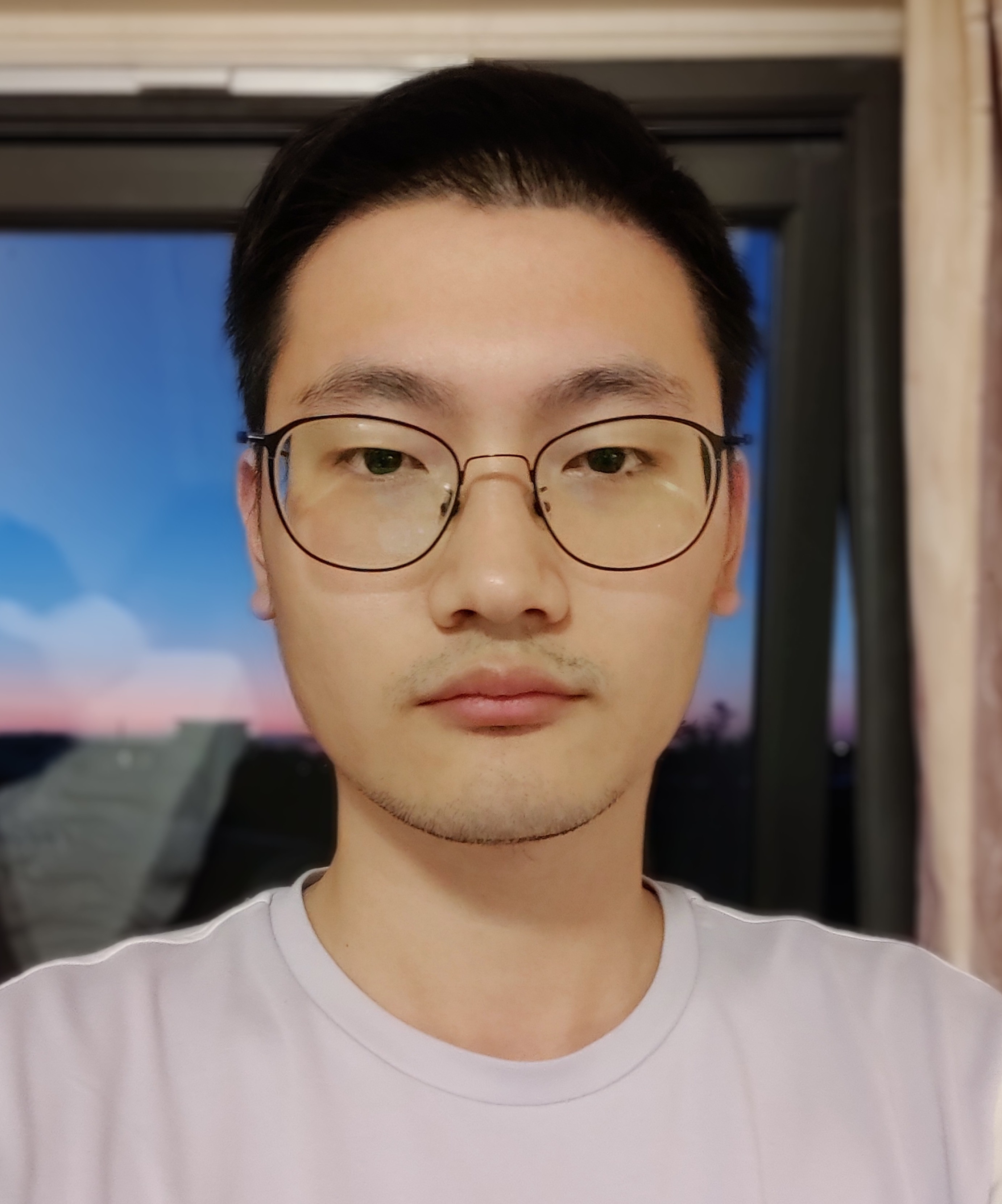}}]{Xuqi Zhu} is currently a Ph.D. student in Computer Science and Electronic Engineering (CSEE) at the University of Essex, UK. He is a member of the Embedded Intelligent Systems Laboratory. His research interests mainly include the custom computing using FPGAs, deep learning for edge devices, and hardware/software co-design.

\end{IEEEbiography}

\begin{IEEEbiography}[{
\includegraphics[width=1.0in,height=1.5in,clip,keepaspectratio]{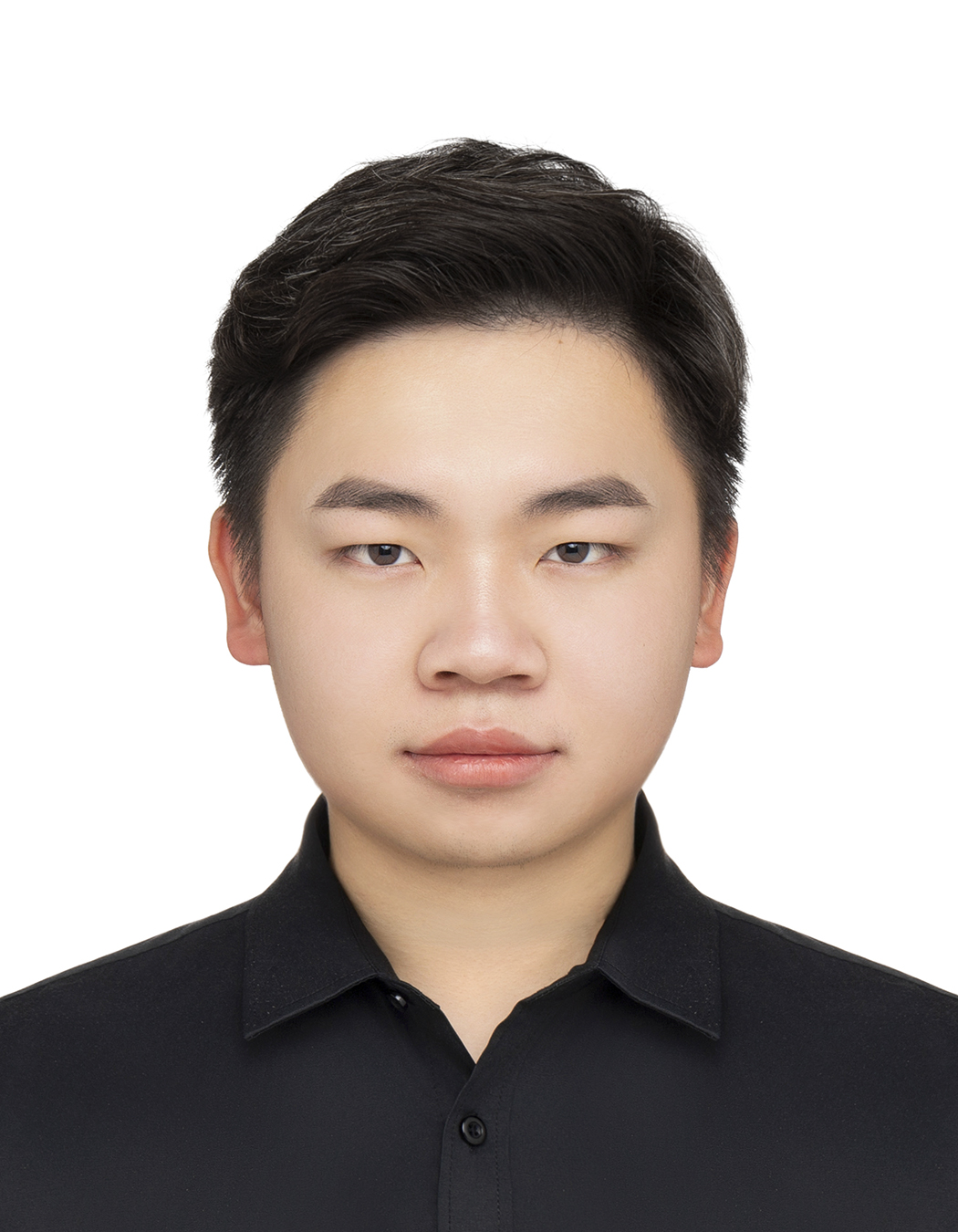}}]{Huaizhi Zhang} is currently pursuing the Ph.D. degree in Computer Science and Electronic Engineering (CSEE) at the University of Essex, UK. He is a member of the Embedded Intelligent Systems Laboratory. His research interests focus on privacy-preserving computation, including but not limited to membership inference attacks (MIA), differential privacy, and their applications in distributed systems. His interdisciplinary work bridges algorithmic privacy guarantees with efficient hardware implementations.

\end{IEEEbiography}

\begin{IEEEbiography}[{
\includegraphics[width=1.0in,height=1.5in,clip,keepaspectratio]{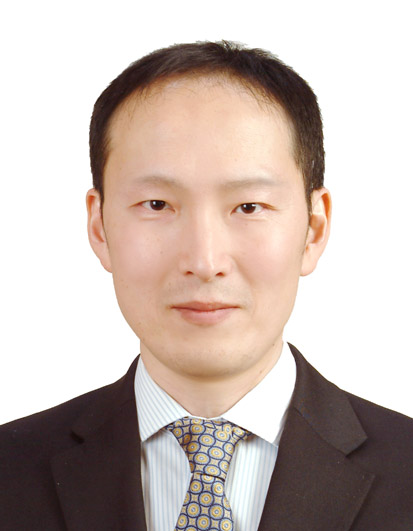}}]{JunKyu Lee} received the Ph.D. degree in computer engineering from The University of Tennessee, Knoxville, TN, USA. He was a postdoctoral researcher with the University of Tennessee, TN, USA, the University of Sydney, NSW, Australia, and Queen’s University Belfast, U.K., respectively. He is currently a Research Fellow with the Institute for Analytics and Data Science, University of Essex, U.K. His research interests include energy-efficient
machine learning and robust machine learning.

\end{IEEEbiography}

\begin{IEEEbiography}[{
\includegraphics[width=1.0in,height=1.5in,clip,keepaspectratio]{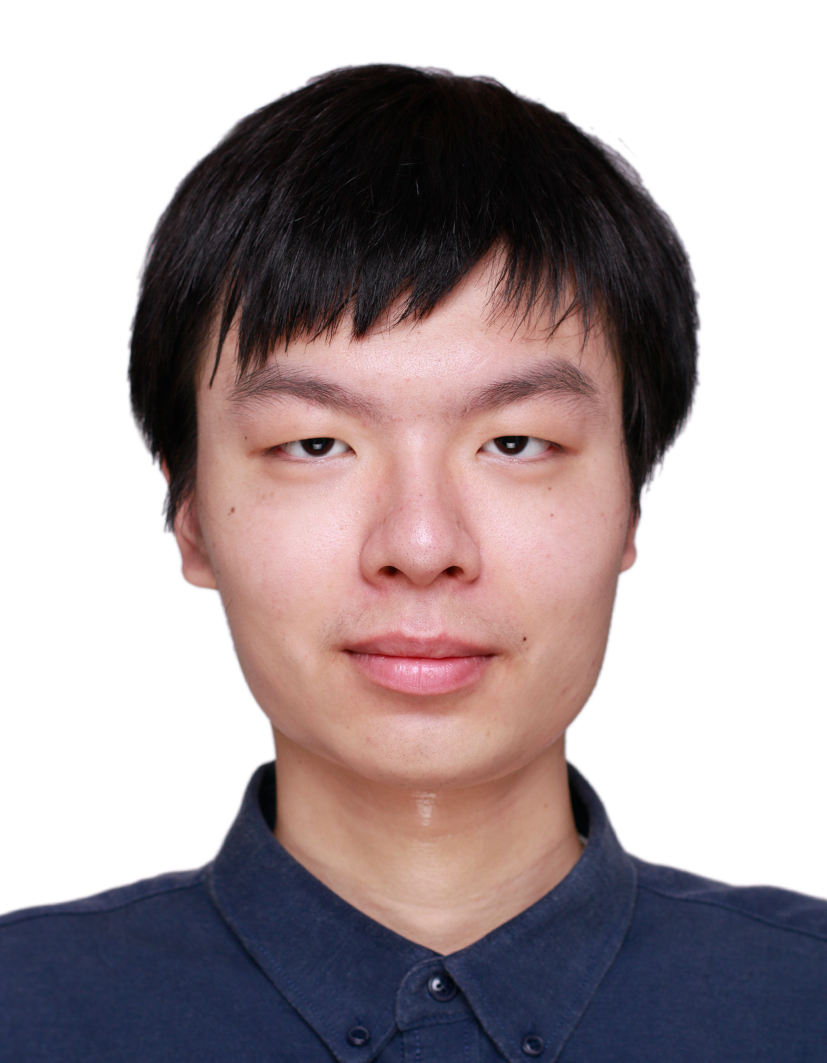}}]{Jiacheng Zhu} is a PhD student from the University of Essex, specializing in hardware accelarator, and possess extensive expertise in FPGA. His experience encompasses a wide range of applications and he is expert in leveraging FPGA technology to innovate and solve complex challenges. Currently, he is assisting the team in prototype development and validation as a research officer of the Embedded and Intelligent Systems Laboratory.

\end{IEEEbiography}

\begin{IEEEbiography}[{
\includegraphics[width=1.0in,height=1.5in,clip,keepaspectratio]{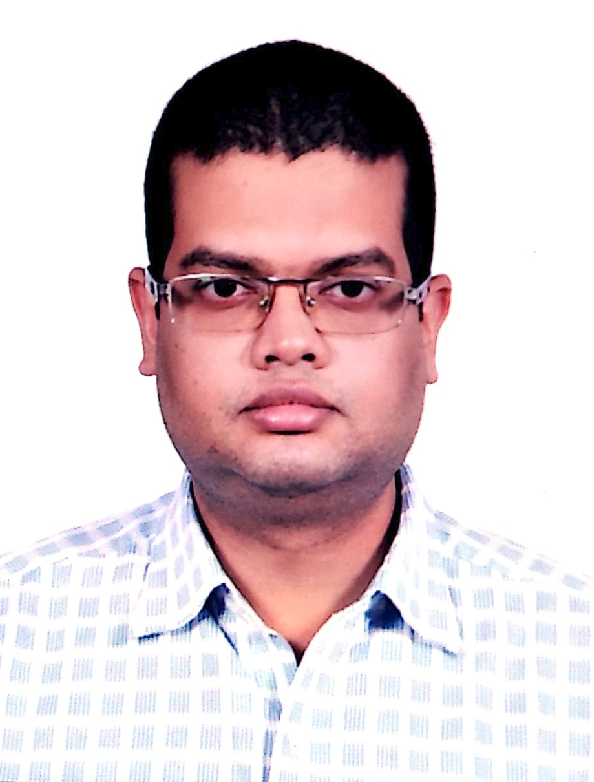}}]{Chandrajit Pal}
is currently associated with the Embedded and Intelligent Systems (EIS) Research Group, University of Essex, UK as a Senior Research Officer. Prior to that, he worked as a National post-doctoral fellow at IIT Hyderabad, India and as an AI Research Engineer at Ceremorphic Inc. 
His research interests mainly include computer vision and signal processing algorithms, custom computing using FPGAs, embedded systems and HW/SW co-design.

\end{IEEEbiography}

\begin{IEEEbiography}[{
\includegraphics[width=1.0in,height=1.5in,clip,keepaspectratio]{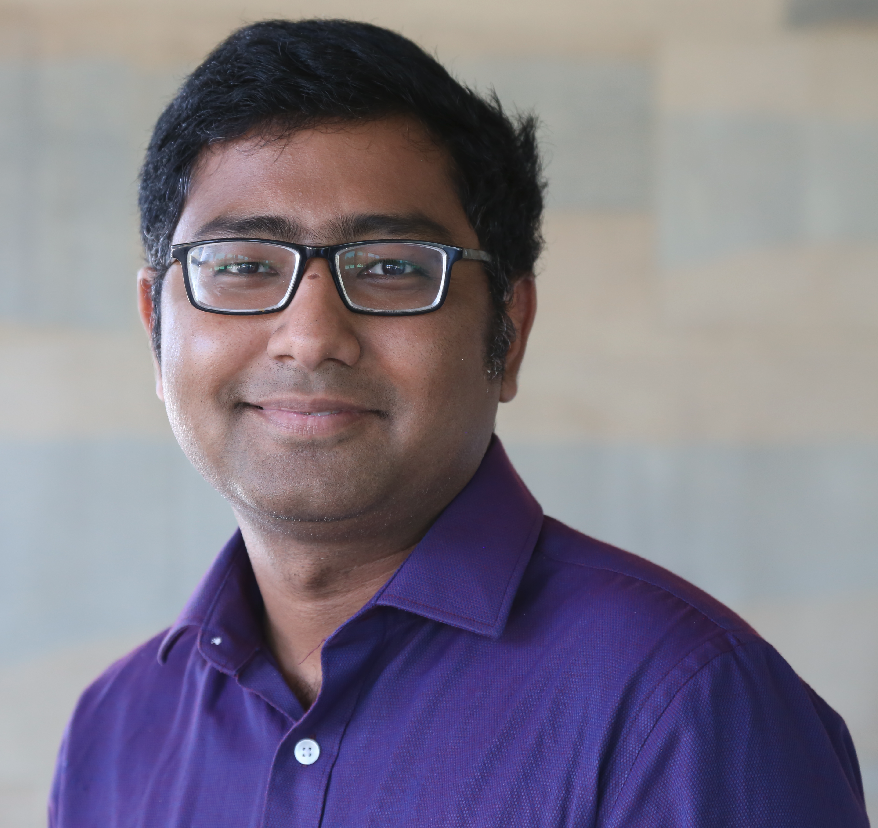}}]{Sangeet Saha}
is currently associated with the Embedded and Intelligent Systems (EIS) Research Group, University of Essex, UK as a Lecturer. Prior to that, he worked as a lecturer at the University of Huddersfield, UK, and Senior research officer (Postdoctoral scholar) at the University of Essex, UK. His current research interests include real-time scheduling, scheduling for reconfigurable computers, real-time and fault-tolerant embedded systems, and cloud computing. 

\end{IEEEbiography}

\begin{IEEEbiography}[{
\includegraphics[width=1.0in,height=1.5in,clip,keepaspectratio]{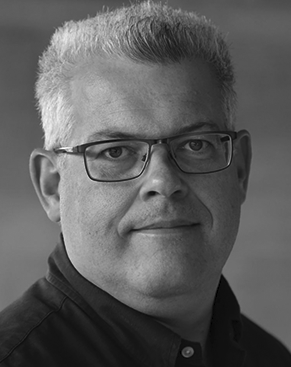}}]{Klaus D. McDonald-Maier}
is currently the head of the Embedded and Intelligent Systems Laboratory and director of research at the University of Essex, Colchester, U.K. He is also the founder of UltraSoC Technologies Ltd., the CEO of Metrarc Ltd., and a visiting professor at the University of Kent. His current research interests include embedded systems and SoC design, security, development support and technology, parallel and energy-efficient architectures, and the application of soft computing and image processing techniques for real-world problems. He is a member of VDE and a fellow of the BCS and IET.

\end{IEEEbiography}

\begin{IEEEbiography}[{
\includegraphics[width=1.0in,height=1.5in,clip,keepaspectratio]{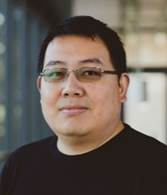}}]{Xiaojun Zhai}
(SM’21) is currently a Reader in the Embedded and Intelligent Systems in School of Computer Science and Electronic Engineering at the University of Essex. His research interests mainly include the design and implementation of machine learning and signal processing algorithms, custom computing using FPGAs, embedded systems and hardware/software co-design. He is a member of ACM and a fellow of IET.

\end{IEEEbiography}

\end{document}